\documentclass[twocolumn]{aastex63}

\usepackage{paralist}
\usepackage{pifont}
\usepackage{bm}
\usepackage{graphicx}

\newcommand{\TRISTAN}{{\bf \textsf{\small Tristan-MP}} }

\newcommand{\alfven}{Alfv\'{e}n }

\usepackage{lineno}

\shorttitle{Electron Acceleration in Quasi-parallel Shocks}
\shortauthors{Gupta, Caprioli, \& Spitkovsky}

\begin{document}

\title{Electron Acceleration at Quasi-parallel Non-relativistic Shocks: A 1D Kinetic Survey}

\correspondingauthor{Siddhartha Gupta}
\email{gsiddhartha@princeton.edu}

\author[0000-0002-1030-8012]{Siddhartha Gupta}
\affiliation{Department of Astrophysical Sciences, Princeton University, 4 Ivy Ln., Princeton, NJ 08544, USA}
\affiliation{Department of Astronomy and Astrophysics, University of Chicago, IL 60637, USA}
\author[0000-0003-0939-8775]{Damiano Caprioli}
\affiliation{Department of Astronomy and Astrophysics, University of Chicago, IL 60637, USA}
\affiliation{Enrico Fermi Institute, The University of Chicago, Chicago, IL 60637, USA}
\author[0000-0001-9179-9054]{Anatoly Spitkovsky}
\affiliation{Department of Astrophysical Sciences, Princeton University, 4 Ivy Ln., Princeton, NJ 08544, USA}

\begin{abstract}
We present a survey of 1D kinetic particle-in-cell simulations of quasi-parallel non-relativistic shocks to identify the environments favorable for electron acceleration.
We explore an unprecedented range of shock speeds $v_{\rm sh}\approx  0.067-0.267\,c$, \alfven Mach numbers $\mathcal{M}_{\rm A} = 5-40$, sonic Mach numbers $\mathcal{M}_{\rm s} = 5-160$, as well as the proton-to-electron mass ratios $m_{\rm i}/m_{\rm e}=16-1836$.
We find that high \alfven Mach number shocks can channel a large fraction of their kinetic energy into nonthermal particles, self-sustaining magnetic turbulence and acceleration to larger and larger energies.
The fraction of injected particles is $\lesssim 0.5\%$ for electrons and $\approx 1\%$ for protons, and the corresponding energy efficiencies are $\lesssim 2\%$ and $\approx 10\%$, respectively.
The extent of the nonthermal tail is sensitive to the \alfven Mach number; when $\mathcal{M}_{\rm A}\lesssim 10$, the nonthermal electron distribution exhibits minimal growth beyond the average momentum of the downstream thermal protons, 
independently of the proton-to-electron mass ratio.
Acceleration is slow for shocks with low sonic Mach numbers, yet nonthermal electrons still achieve momenta exceeding the downstream thermal proton momentum when the shock \alfven Mach number is large enough.
We provide simulation-based parametrizations of the transition from thermal to nonthermal distribution in the downstream (found at a momentum around $p_{\rm i,e}/m_{\rm i}v_{\rm sh} \approx 3\sqrt{m_{\rm i,e}/m_{\rm i}}$), as well as the ratio of nonthermal electron to proton number density.
The results are applicable to many different environments and are important for modeling shock-powered nonthermal radiation.
\end{abstract}


\section{Introduction} \label{sec:intro}
%
One of the biggest open questions in high-energy astrophysics and space physics is how energetic electrons are accelerated in collisionless shocks, where Coulomb interactions are negligible.
While the production of these particles in different sources such as the Earth's bow shock \citep[e.g.,][]{liu+19}, shocks in supernova remnants \citep[e.g.,][]{SN1006HESS} and other galactic \citep[e.g.,][]{su+10} or extra-galactic environments \citep[e.g.,][]{vanweeren+10} is inferred from synchrotron radio, X-rays, and inverse-Compton $\gamma$-rays emission,
no available theory can predict the dependence of electron acceleration efficiency on parameters such as the shock velocity, the upstream plasma temperature, and the background magnetic field strength and inclination relative to the shock normal.

The current theory of particle acceleration at space/astrophysical collisionless shocks is based on a kind of Fermi mechanism called diffusive shock acceleration (DSA) \citep[][]{Fermi49,krymskii77,axford+77p,bell78a,blandford+78}, which naturally predicts power-law spectra of nonthermal (NT) particles \citep[see][for reviews]{drury83, marcowith+20, caprioli23}.
However, which processes are responsible for electron injection into DSA is not well understood. 
The main issue is that, in order to see the upstream and downstream as converging flows while  neglecting the shock structure, the particle Larmor radius is generally expected to be larger than the shock transition, which varies between the proton skin depth to the proton Larmor radius \citep[e.g.,][]{treumann09};
both scales are larger than the Larmor radius of thermal electrons. 
Furthermore, there remains uncertainty regarding whether such prerequisites, i.e., the Larmor radius being larger than the shock transition, are truly essential for injection into DSA \citep{schwartz+83,burgess+84,caprioli+15}.

In the case of protons, observations and kinetic simulations suggest that $10-20\%$ of upstream bulk kinetic energy goes into NT protons, consistent with the efficiency required to account for the energy budget in Galactic cosmic rays (CRs) \citep{caprioli+14a,caprioli+17,gupta+20}. 
Proton injection from the thermal bath is favored in quasi-parallel regimes, where the inclination of the magnetic field relative to the shock normal is $\theta_{\rm Bn}\lesssim 50^{\rm o}$ \citep[][]{SN1006HESS,caprioli+14a}, or generally in weakly-magnetized, non-relativistic shocks \citep{orusa+23}.
However, both quasi-parallel and quasi-perpendicular/oblique shocks ($\theta_{\rm Bn}\gtrsim 50^{\rm o}$) are often associated with energetic electrons \citep[]{SN1006HESS,amano+07,amano+09a,guo+14a, park+15, liu+19, winner+20, xu+20, shalaby+22, morris+22}.
This suggests that electron injection may work differently from proton injection, with different scaling dependence on environmental parameters.

Quantifying the fraction of bulk kinetic energy that goes to energetic electrons, or the electron acceleration efficiency, 
through observations is also challenging.
In general, to explain synchrotron emission from radio to X-rays, and inverse-Compton emission in the $\gamma$-rays, even in the same class of sources, the NT electron fraction in collisionless shocks may vary by orders of magnitude \citep[e.g.,][]{chevalier+06,panaitescu+01,sarbadhicary+17,merten+17,reynolds+21}.
Finally, a dependence on the shock speed is also likely, since for relativistic shocks the difference between electrons and protons is significantly reduced, as attested by both simulations \citep{sironi+09, sironi+11, sironi+13, crumley+19} and modeling of gamma-ray burst afterglows \citep[e.g.,][]{piran04}.

To quantify electron acceleration one needs to account for the non-linear interplay between thermal/NT particles and electromagnetic (EM) fields self-consistently, which requires first-principles kinetic approaches, typically particle-in-cell (PIC) simulations.
In past decades, several studies found compelling evidence of electron acceleration at non-relativistic shocks \citep[e.g.,][]{shimada+00,riquelme+11,sironi+13,guo+14a,guo+14b,park+15, crumley+19,bohdan+19a,xu+20,shalaby+22,morris+22}.
A few works also reported the development of electron power-law tail in both non-relativistic and (trans-)relativistic regimes \citep[e.g.,][]{sironi+13,kato15,park+15,crumley+19,xu+20,arbutina+21, kumar+21,shalaby+22,morris+22, grassi+23,zekovic+24}.
However, a systematic measurement of electron acceleration efficiency for different shock parameters remains to be completed.

In this work, we present a detailed survey of kinetic shock simulations by focusing on quasi-parallel shocks, de facto extending the pioneering work of \citet{park+15}, which found a clear signature of electron and proton DSA at non-relativistic quasi-parallel shocks.
We cover a range of shock parameters, such as speed and magnetization, and also explore the role of the plasma $\beta$ (i.e., the ratio of thermal pressure to magnetic pressure) in typical interstellar ($\beta\sim 1$) and intracluster ($\beta\gg 10$) conditions.

The key questions addressed in this work are
\begin{compactitem}
\setlength\itemsep{0.07em}
    \item Do electron injection and acceleration depend on the shock speed?
    \item How do they depend on the shock \alfven and sonic Mach numbers ($\mathcal{M}_{\rm A}$ and $\mathcal{M}_{\rm s}$)?
    \item How can we scale the results from reduced proton-to-electron mass ratios that are normally used in PIC simulations to realistic values? 
\end{compactitem}
We provide several observational parameters, such as the fractions of energy channeled into NT particles, thermal plasma, and magnetic fields, and the electron-to-proton temperature ratios behind the shock.
Our results suggest that high-$\mathcal{M_A}$ quasi-parallel shocks accelerate both electrons and protons efficiently.

We present our simulation setup in \S\ref{sec:simusetup} and discuss the basic properties of the shock structure in \S\ref{sec:shockstruc}.
The self-generated magnetic fields produced by accelerated protons are presented in \S \ref{sec:emprofile}, while in \S \ref{sec:accsig} we discuss how acceleration occurs in different environments.
We also provide several useful observational diagnostics in \S\ref{sec:scalings}, compare our results with previous studies in \S\ref{sec:compre}, and finally summarize our findings in \S\ref{sec:conclusion}.
%
\section{Shock setup}\label{sec:simusetup}
\begin{table}
\begin{footnotesize}
\begin{center}
\begin{tabular}{l |  c c c c c c | l}
   \hline\hline
   Runs  & $\frac{v_{\rm pt}}{c}$ & $m_{\rm R}$  & $\mathcal{M}_{\rm A}$  & $\mathcal{M}_{\rm s}$  & $\mathcal{M}_{\rm se}$& $\beta$  & Figures \\ 
    \hline
    $\mathcal{A}1$            & $0.05$   & $100$   &  $20$  & $40$ & $4$  & $1$    &   \ref{fig:x-vx-vp} and \ref{fig:spec_vp} \\
    \hline
    $\mathcal{B}1^{\star}$    & $0.1$&   $100$   &  $20$  & $40$  & $4$   & $1$    &  \ref{fig:x-vx-vp} and \ref{fig:spec_vp}  \\
    $\mathcal{B}1_{\rm II}$   & $0.1$&   $16$    &  $20$  & $40$  & $10$  & $1$    & \ref{fig:spec-mr} \\
    $\mathcal{B}1_{\rm III}$  & $0.1$&   $400$   &  $20$  & $40$  & $2$   & $1$    &  \ref{fig:spec-mr} \\
     $\mathcal{B}1_{\rm IV}$  & $0.1$&   $1836$  &  $20$  & $40$  & $0.9$ & $1$    & \ref{fig:spec-mr} \\
    $\mathcal{B}2$            & $0.1$&   $100$   &  $5$   & $40$  & $4$   & $0.06$ &  \ref{fig:spec-Ma-Ms-MaMs} \\
    $\mathcal{B}2_{\rm II}$   & $0.1$&   $1836$  &  $5$   & $40$  & $0.9$ & $0.06$ & \ref{fig:spec-mr}  \\
    $\mathcal{B}3$            & $0.1$&   $100$   &  $10$  & $40$  & $4.$  & $1$    &  \ref{fig:spec-Ma-Ms-MaMs}  \\
     $\mathcal{B}4$           & $0.1$&   $100$   &  $20$  & $5$  & $0.5$   & $64$  &  \ref{fig:spec-Ma-Ms-MaMs} \\
     $\mathcal{B}5$           & $0.1$&   $100$   &  $20$  & $10$  & $1$   & $16$   & \ref{fig:spec-Ma-Ms-MaMs} \\
     $\mathcal{B}6$           & $0.1$&   $100$   &  $20$  & $160$ & $16$  & $0.06$ & \ref{fig:spec-Ma-Ms-MaMs} \\
     $\mathcal{B}7$           & $0.1$&   $100$   &  $5$  & $10$  & $1$    & $1$    &   \ref{fig:spec-Ma-Ms-MaMs} \\
    $\mathcal{B}8$            & $0.1$&   $100$   &  $10$  & $20$  & $2$   & $1$    &  \ref{fig:spec-Ma-Ms-MaMs} \\
      $\mathcal{B}9$          & $0.1$&   $100$   &  $30$  & $60$  & $6$   & $1$    &  \ref{fig:spec-Ma-Ms-MaMs} \\
     $\mathcal{B}10$          & $0.1$&   $100$   &  $40$  & $80$  & $8$   & $1$    &   \ref{fig:spec-Ma-Ms-MaMs} \\
     \hline 
    $\mathcal{C}1$             & $0.2$ &  $100$   & $20$   & $40$  & $4$  & $1$    & \ref{fig:x-vx-vp} and \ref{fig:spec_vp}\\
    \hline
\end{tabular}
\end{center}
\centering
\caption{
Simulation parameters:
$v_{\rm pt}$ is the piston speed, $m_{\rm R}$ is the proton-to-electron mass ratio, $\mathcal{M}_{\rm A}$ is the Alfv\'{e}n Mach number, $\mathcal{M}_{\rm s}$ and $\mathcal{M}_{\rm se}$ denote the sonic Mach number of protons and electrons, respectively.
$\beta\equiv 4(\mathcal{M}_{\rm A}/\mathcal{M}_{\rm s})^2$ is the ratio of thermal pressure to magnetic pressure.
The rightmost column indicates in which figure the particle spectra are shown. 
The symbol $\star$ marks the fiducial run.}\label{tab:simpara}
\end{footnotesize}
\end{table}
We perform kinetic shock simulations using the EM PIC code \TRISTAN (\citealt{spitkovsky05}).
We consider a quasi-1D geometry (five cells along the transverse $y$-axis\footnote{With five $y$-cells, the effective number of particles per $d_{\rm e}$ is increased by a factor of five compared to the traditional one $y$-cell choice. 
This approach improves particle statistics while keeping the EM profiles similar.}) with grid and time spacing fixed to $\Delta x = d_{\rm e}/10$ and $\Delta t = 0.045\, \omega_{\rm pe}^{-1}$ respectively, where $d_{\rm e}=c/\omega_{\rm pe}$ is the electron skin depth, and $\omega_{\rm pe}=\sqrt{4\pi n_{\rm 0}\,e^2/m_{\rm e}}$ is the electron plasma frequency; $n_{\rm 0}$ is the plasma density, while $e$ and $m_{\rm e}$ are the electron charge and mass.
Each cell is filled with an electron-proton plasma with $200$ particles per cell per species.
Electrons and protons are initialized with a Maxwell-Boltzmann distribution in thermal equilibrium (i.e., their initial temperatures $T_{\rm 0i}=T_{\rm 0e}$).

The shock is launched using the left boundary of the computation domain as a moving reflecting wall (hereafter, the `piston'). 
The piston velocity, $v_{\rm pt}$, the initial orientation of the magnetic field (${\bf B}_{\rm 0}$) relative to the shock/piston normal, $\theta_{\rm Bn}$, the thermal speed of protons and electrons, $v_{\rm thi,e}=\sqrt{k_{\rm B} T_{\rm 0i,e}/m_{\rm i,e}}$, and the Alfv\'{e}n speed $v_{\rm A}\equiv B_{\rm 0}/\sqrt{4\pi m_{\rm i} n_{\rm 0}}$ are free parameters, which are chosen to study a specific shock environment.
As the piston moves along the $x$-axis with $v_{\rm pt}\gg v_{\rm A}$, it sweeps up plasma and produces a shock.
We focus on quasi-parallel shocks and initialize ${\bf B}_{\rm 0}$ in the $x-y$ plane with $\theta_{\rm Bn}=\cos^{-1}(B_{\rm x}/|{\bf B}|)=30^{\rm o}$ and $\phi_{\rm Bn}=\tan^{-1}(B_{\rm y}/B_{\rm z})=90^{\rm o}$.
For both particles and fields, the right (left) boundary of the computational domain is open (reflecting); other boundaries are periodic.
 
Shocks in our simulations are characterized by three parameters: shock speed $v_{\rm sh}$, Alfv\'{e}nic Mach number $\mathcal{M}_{\rm A}=v_{\rm sh}/v_{\rm A}$, and sonic Mach number $\mathcal{M}_{\rm s}=v_{\rm sh}/v_{\rm th,i}$.
To infer the shock speed $v_{\rm sh}$ from the piston speed, we assume $v_{\rm sh}=v_{\rm pt}\mathcal{R}/(\mathcal{R}-1)$ where the density compression ratio is $\mathcal{R}=\rho_{\rm 2}/\rho_{\rm 1}=4$ (subscripts $1$ and $2$ refer to the upstream and downstream quantities, respectively).
However, we note that the actual shock Mach numbers (as calculated in the shock frame) can vary by $\approx \pm 10\%$ depending on the compression ratio $\mathcal{R}$ found in self-consistent simulations \citep[e.g.,][]{caprioli+14a,haggerty+20}.
The displacement of the shock relative to the piston is
\begin{equation}\label{eq:xsh}
    x_{\rm sh} = \left(v_{\rm sh} - v_{\rm pt}\right) t= \frac{\mathcal{M}_{\rm A}}{\mathcal{R}} \frac{t}{\omega_{\rm ci}^{-1}}\,d_{\rm i}\, ,
\end{equation}
where $d_{\rm i}=\sqrt{m}_{\rm R}\,d_{\rm e}$ is the proton skin depth, $m_{\rm R} =m_{\rm i}/m_{\rm e}$ is the proton-to-electron mass ratio, and $\omega_{\rm ci}\equiv v_{\rm A}/d_{\rm i}$ is the proton cyclotron frequency in the initial magnetic field.

To make our runs computationally tractable, we use a reduced mass ratio $m_{\rm R}=100$.
Since the reduced mass ratio changes the effective electron sonic Mach number $\mathcal{M}_{\rm se}=\mathcal{M}_{\rm s}/\sqrt{m}_{\rm R}$, we have also explored $m_{\rm R}=16,40,$ and $1836$ to quantify the differences.

The setup described above is similar to the previous PIC campaigns \citep[e.g.,][]{sironi+11,guo+14a,park+15,crumley+19} with the exception that our simulations are in the upstream rest frame, similar to \citet{xu+20}.
After a careful investigation, we found that when the Debye length ($\lambda_{\rm D}\equiv v_{\rm the}/\omega_{\rm pe}$) is marginally resolved or under-resolved, the upstream rest frame is an excellent choice to avoid numerical heating for the parameter space used in this paper. 
We also filter the current with 32 passes of a digital filter (with 0.25-0.5-0.25 weighting) to reduce numerical noise \citep[e.g.,][]{shalaby+17,arbutina+21}.
We tested the impact of different numbers of filter passes and found the results to be insensitive to them.
To capture the back-reaction of the highest energy particles, we use a receding right boundary that enlarges the size of the domain along the $x$-axis with time to $\sim 10^4\,d_{\rm i}$ for a run time of $300\,\omega_{\rm ci}^{-1}$.
The key parameters of all runs are detailed in Table \ref{tab:simpara}.
\section{1D quasi-parallel shocks}\label{sec:shockstruc}
\begin{figure}
    \centering
    \includegraphics[width=3.4in]{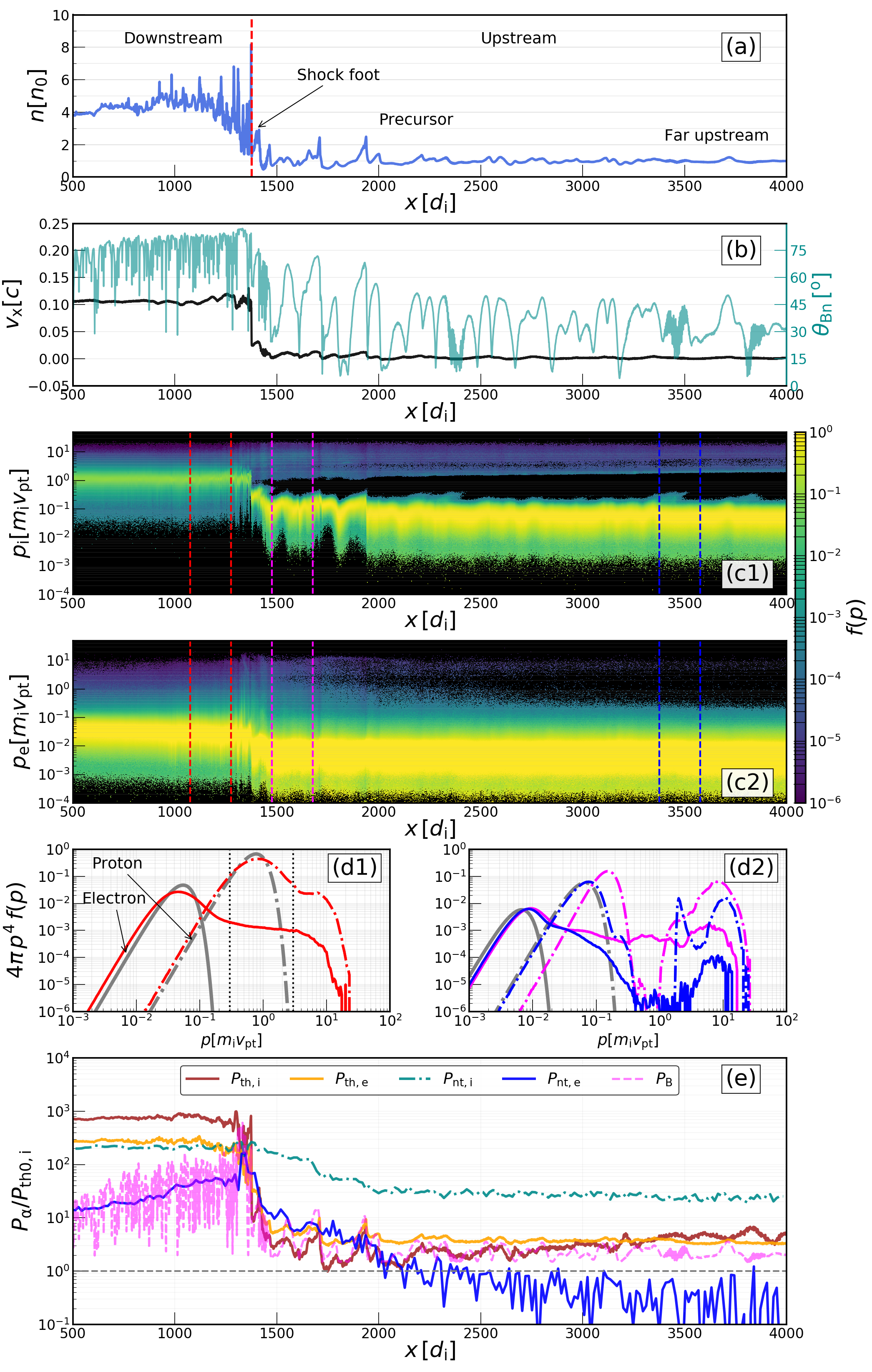} 
    \caption{
    Snapshot of different profiles at $t=275\,\omega_{\rm ci}^{-1}$ obtained from our benchmark simulation  ($v_{\rm pt}/c=0.1$, $\mathcal{M}_{\rm A}=20$, and $\mathcal{M}_{\rm s}=40$, the run $\mathcal{B}1$ in Table \ref{tab:simpara}).
    Panel (a) displays the spatial distribution of the proton number density (normalized to the far upstream density $n_{\rm 0}$), panel (b) $x$-component of the plasma velocity (left axis) and shock inclination (right axis).
    Panels (c1) and (c2) show the $x-|p|$ phase-space distribution of protons and electrons, respectively.
    Panels (d1) and (d2) represent the spectra of the three regions marked in panels (c1) and (c2), using the same color scheme for the lines.
    The dotted vertical lines in panel (d1) mark the momentum identified as the boundary between thermal and NT distributions.
    Panel (e) shows the different pressure components (thermal, NT, and magnetic field) normalized to the initial thermal pressure of protons ($P_{\rm th 0,i}$).
    }\label{fig:shockstruc}
\end{figure}

%
Before presenting the results from the survey, we take a moment to show some crucial shock profiles from our benchmark simulation $\mathcal{B}1$ (Table \ref{tab:simpara}), where $v_{\rm pt}/c=0.1$, $\mathcal{M}_{\rm A}\equiv \mathcal{M}_{\rm s}/2=20$, and $m_{\rm R}=100$ (as used in \citealt{park+15}, but in the upstream frame) in Figure \ref{fig:shockstruc}.
\subsection{Density}\label{subsec:density}
%
Figure \ref{fig:shockstruc}(a) displays a snapshot of the plasma (thermal + NT) density profile at $t=275\,\omega_{\rm ci}^{-1}$, when the shock has already achieved a quasi-stationary structure.
The horizontal axis shows the distance from the location of the piston and the red line marks the location of the shock, $x_{\rm sh} \approx 1400\,d_{\rm i}$ (also see Equation \ref{eq:xsh}).
While the density immediately behind the shock is slightly higher than four, the spatially averaged compression is consistent with the Rankine-Hugoniot shock jump condition \citep[e.g.,][]{shu92}.
\subsection{Velocity and the Inclination of Magnetic Field}\label{subsec:profilevelo}
%
The black curve in Figure \ref{fig:shockstruc}(b) shows the $x$-component of the bulk plasma velocity ($v_{\rm x}$).
Since our simulations are in the upstream rest frame, $v_{\rm x}$ in the downstream is approximately the same as the piston speed.
The cyan curve represents the local direction of the magnetic field, which is significantly different from the far upstream value of $\theta_{\rm Bn}=30^{\rm o}$, due to the proton-driven streaming instabilities that develop magnetic fluctuations  \citep[e.g.,][]{bell05,amato+09,bell+13,caprioli+14b}.
These modifications regulate both proton and electron injection into DSA.
\subsection{Phase space}\label{subsec:profilexp}
%
Figures \ref{fig:shockstruc}(c1) and (c2) display the $x-|p|$ phase-space distributions of protons and electrons, respectively.
The upstream proton phase-space shows two distinct populations -- one is thermal and the other encompasses energetic/NT protons.
Besides these populations, the electron phase-space in the upstream contains an additional population, namely, suprathermal (ST) electrons (with $p/m_{\rm i}v_{\rm pt}\sim 10^{-2}$).
This population comprises both electrons reflected by the shock and electrons generated locally to neutralize the total current in the plasma, as detailed in \citet{gupta+24a}.
\subsection{Spectra}\label{subsec:spec}
%
Figures \ref{fig:shockstruc}(d1) and (d2) show the momentum distribution of electrons and protons (solid and dash-dotted curves),  for the downstream (red) and upstream (purple and blue) regions marked in the figures above.
In both panels, the distributions are multiplied by $p^4$ to facilitate the comparison with the standard DSA prediction, $f(p)\propto p^{-4}$.
The grey curves in Figures \ref{fig:shockstruc}(d1) and (d2)  show the thermal Maxwellian distribution, i.e.,
\begin{eqnarray}\label{eq:dnp}
 4\pi p^4 f_{\rm th}(p) & = & 4\pi\,p^4\,\frac{ 1}{(2\pi\,m k_{\rm B} T)^{3/2}}\exp\left[-\frac{p^2}{2\,m k_{\rm B}T}\right].
\end{eqnarray}  
Throughout the paper we normalize momenta by the piston momentum, $m_{\rm i}v_{\rm pt}$, as it makes the resulting spectra independent of the shock speed, as discussed below.
\subsubsection{Thermal Distributions} 
The peak of the downstream thermal distribution for each species $\alpha\in i,e$ in the $4\pi p^4 f(p)$ representation is expected at $2\sqrt{m_{\rm \alpha} k_{\rm B} T_{\rm  \alpha}}/m_{\rm i}v_{\rm pt} =0.82\sqrt{m_{\rm  \alpha}/m_{\rm i}}$, when the postshock gas follows the Rankine-Hugoniot strong shock condition that gives $k_{\rm B}T_{\rm \alpha}=\mu_{\rm \alpha} (\gamma_{\rm th}-1)m_{\rm i}v_{\rm pt}^2/2$ (with $\gamma_{\rm th}=5/3$), and equipartition between electrons and protons is assumed (i.e., $\mu_{\rm i,e}\equiv T_{\rm i,e}/(T_{\rm i}+T_{\rm e}) =1/2$).
Note that in these units the downstream proton spectrum depends neither on the shock speed nor on the mass ratio, and the peak of the electron distribution relative to the protons' depends on $\sqrt{m_{\rm R}}$ and  $(T_{\rm e}/T_{\rm i})$, where $T_{\rm e}/T_{\rm i}$ is typically close to $\sim 1/2$ in the downstream.

Similarly, the upstream thermal momentum distribution peaks at $2\sqrt{m_{\rm\alpha} k_{\rm B} T_{\rm \alpha}}/m_{\rm i}v_{\rm pt} = 2.67 \sqrt{m_{\rm \alpha}/m_{\rm i}}/\mathcal{M}_{\rm s}$, i.e., it depends only on $\mathcal{M}_{\rm s}$ and $\sqrt{m_{\rm \alpha}/m_{\rm i}}$,
which can be seen from Figure \ref{fig:shockstruc}(d2) using $\mathcal{M}_{\rm s}=40$; 
however, it appears to be slightly hotter than that specified at $t=0$ due to the back-reaction from energetic particles \citep{gupta+24a}.
Note that particles achieve a thermal distribution in the downstream in a collisionless way, i.e., due to interactions with EM fields rather than Coulomb collisions, with $T_{\rm e}$ only marginally smaller than $T_{\rm i}$ \citep[e.g.,][]{vanthieghem+24}.
\subsubsection{Nonthermal Distributions}\label{subsubsec:NTdis}
%
Above the thermal peaks, power-law distributions of NT particles arise for both species.
While quantifying the precise transition point between the thermal and the nonthermal particle is generally challenging,
Figure \ref{fig:shockstruc}(d1) suggests that the transition occurs at $p/m_{\rm i} v_{\rm pt}\approx 0.3$ and $3$ respectively, as shown by the vertical dotted lines, which correspond to $p_{\rm 0,\alpha}\equiv 5\,p_{\rm th,\alpha}$, where $p_{\rm th,\alpha}=\sqrt{2\,m_{\rm \alpha}\,k_{\rm B} T_{\rm \alpha}}\equiv 0.58 \sqrt{2\,\mu_{\rm \alpha}\,m_{\alpha} m_{\rm i}}\, v_{\rm pt}$ is the peak momentum of the downstream thermal distribution for the respective species ($\alpha \in i,e$).
These lower ends of the NT distributions can be rephrased as
\begin{equation}\label{eq:p0alpha}
        p_{\rm 0,\alpha} \simeq 3 \sqrt{2\mu_{\rm \alpha} \frac{m_{\rm \alpha}}{m_{\rm i}}} m_{\rm i}v_{\rm pt}
\end{equation}
where $\mu_{\rm \alpha}\equiv T_{\rm \alpha}/(T_{\rm i}+T_{\rm e})\sim 1/2$ if the downstream protons and electrons are in thermal equilibrium.
The slope of the downstream NT distributions for both electrons and protons is very close to $4$, consistent with the DSA prediction for strong shocks.

The upstream spectra shown in Figure \ref{fig:shockstruc}(d2) are complex, and their shapes depend on the distance from the shock.
Near the shock, where particles diffuse (shock precursor), spectra may exhibit rapid fluctuations depending on the magnetic structures, but are in general agreement with DSA expectations \citep[e.g.,][]{caprioli+10a}.
Further upstream, i.e., beyond the shock precursor (blue curves in Figure \ref{fig:shockstruc}(d2)), the top-hat distribution primarily represents the escaping populations of both energetic electrons and protons.
Besides such populations, the electron distribution contains shock-reflected electrons and the locally produced suprathermal electrons that compensate the current in NT protons \citep[][]{gupta+24a}.

\begin{figure*}[ht!]
\centering
\includegraphics[width=6.5in]{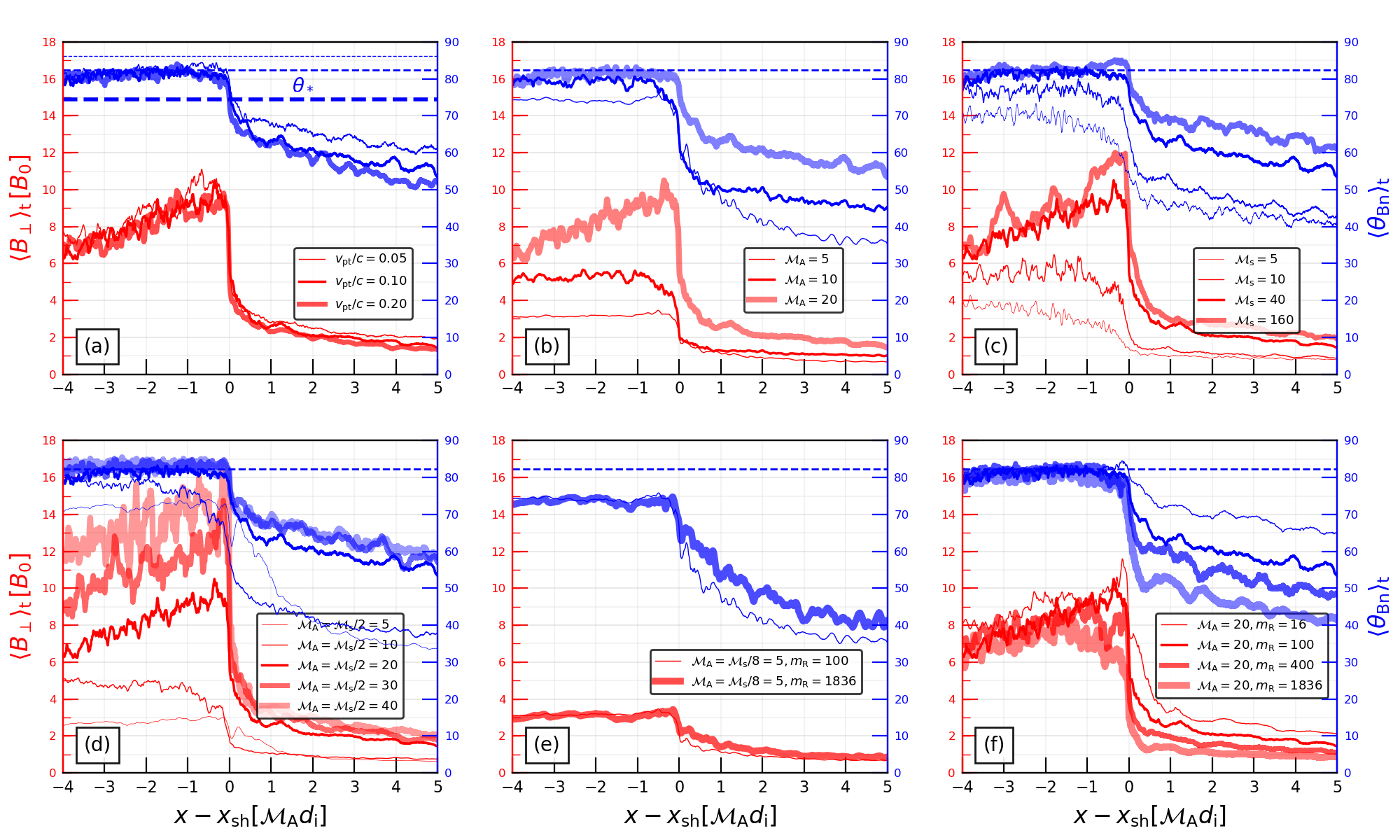}
\caption{Time averaged (between $100-150\,\omega_{\rm ci}^{-1}$) $B_{\rm \perp}=\sqrt{B_{\rm y}^2+B_{\rm z}^2}$ and $\theta_{\rm Bn}=\cos^{-1}(B_{\rm x}/|{\bf B}|)$ for different shock parameters.
The left and right axes represent $\langle B_{\rm \perp}\rangle _{\rm t}$ and $\langle \theta_{\rm Bn}\rangle_{\rm t}$, respectively.
The horizontal axis in all panels is normalized to $\mathcal{M}_{\rm A}d_{\rm i}$.
The dashed horizontal line(s) in each panel shows the $\theta_{*}=\cos^{-1}(v_{\rm sh}/c)$ above which the shock becomes superluminal.
Panels (a)--(f) represent different runs where $v_{\rm pt}/c=0.1,\,\mathcal{M}_{\rm A}=20,\mathcal{M}_{\rm s}=40,$ and $m_{\rm R}=100$, unless otherwise mentioned.
}
\label{fig:magtheta}
\end{figure*}
\subsection{Thermal and NT Pressures}\label{subsec:pressure}
%
To obtain the profile of the pressure in thermal and NT populations, we assume that any particle below a momentum of $p_{\rm 0,\alpha}$ (Equation \ref{eq:p0alpha}) is thermal and above $p_{\rm 0,\alpha}$ is NT, as discussed in \S \ref{subsubsec:NTdis}.
Such a threshold is set by the transition observed in the downstream spectra, i.e., the power-law tail for both protons and electrons starts at $p_{\rm i,e}/m_{\rm i}v_{\rm pt}\approx 3\sqrt{m_{\rm i,e}/m_{\rm i}}$ (Figure \ref{fig:shockstruc}(d1)).
We use the same momentum threshold also upstream, acknowledging that this may include suprathermal particles in the thermal pressure budget (Figure \ref{fig:shockstruc}(d2)).

The pressures are estimated locally in the rest frame of the plasma using
\begin{equation}
P_{\rm rs,\alpha} \equiv n_{\rm \alpha}\int \gamma(p)\, u_{\rm r}(p)\, u_{\rm s}(p)\, f(p) d^3p\ ,
\end{equation}
where $r,s\in x,y,z$, $u_{\rm r}$ is the $r$-component of a particle's $3$-velocity in the comoving frame, and $n_{\rm \alpha}$ is the number density at a given location for the species $\alpha$.
Figure \ref{fig:shockstruc}(e) displays the pressure profiles of the thermal particles  ($P_{\rm th,i}$: protons and $P_{\rm th, e}$: electrons), the NT energetic particles ($P_{\rm nt,i}$: protons and $P_{\rm nt, e}$: electrons),
along with the magnetic pressure ($P_{\rm B}=B^2/8\pi$), all pressures are normalized to the initial thermal proton pressure $P_{\rm th 0,i}$.

The proton thermal pressure (brown line) experiences a jump of $P_{\rm th, i}/P_{\rm th 0,i}\approx 700$, which is approximately $40\%$ smaller than the one predicted for a hydrodynamic shock $(2 \mathcal{M}_{\rm s}^2-(\gamma_{\rm th}-1))/(\gamma_{\rm th}+1)\,\simeq 1200$. 
A smaller jump is expected since the shock kinetic energy is also processed into thermal electrons, magnetic energy, and energetic populations, which are not included in the single-fluid hydrodynamic description.
The electron thermal pressure (orange line) is smaller than $P_{\rm th, i}$ by a factor of $3$, indicating the electron-to-proton temperature ratio $T_{\rm e}/T_{\rm i}\sim 0.3$, i.e., downstream electrons are slightly cooler than protons.
Interestingly, the magnetic pressure is lower than the thermal and NT proton pressures, and it is correlated with the NT electron pressure.
This suggests that the dynamics of NT electrons may be connected to that of the self-generated magnetic fields.

The values of $P_{\rm th,i}$ and $P_{\rm th, e}$ just upstream of the shock are larger than $P_{\rm th 0,i}$ by a factor $\sim 2$.
The magnetic pressure (magenta line) is also a factor of $\sim 2$ larger than its initial value, 
i.e., thermal and magnetic pressure remain close to far-upstream equipartition values due to the back-reaction from streaming instabilities \citep[e.g.,][]{bell04,riquelme+09,gupta+21p,zacharegkas+24}.
The green and blue lines representing the pressure in NT protons and electrons show the exponential fall-off shape, which is characteristic of diffusing particles \citep[e.g.,][]{caprioli+14a}.

Below we present the results from our survey simulations (Table \ref{tab:simpara}): first, the self-generated EM fields (\S\ref{sec:emprofile}), followed by the spectra of electrons and protons (\S\ref{sec:accsig}), and finally, the physical scalings for different observational parameters (\S \ref{sec:scalings}).
\section{Electromagnetic Profiles}\label{sec:emprofile}
%
Quasi-parallel shocks naturally develop back-streaming particles, which drive instabilities and modify the magnetic field topology and $\theta_{\rm Bn}$, as we show in Figure \ref{fig:magtheta}.
Such modifications are pivotal to regulate the efficiency in producing NT particles  \citep[e.g.,][]{bell04,caprioli+14b,park+15,gupta+24a}.

\begin{figure*}[ht!]
\centering
\includegraphics[width=6.4in]{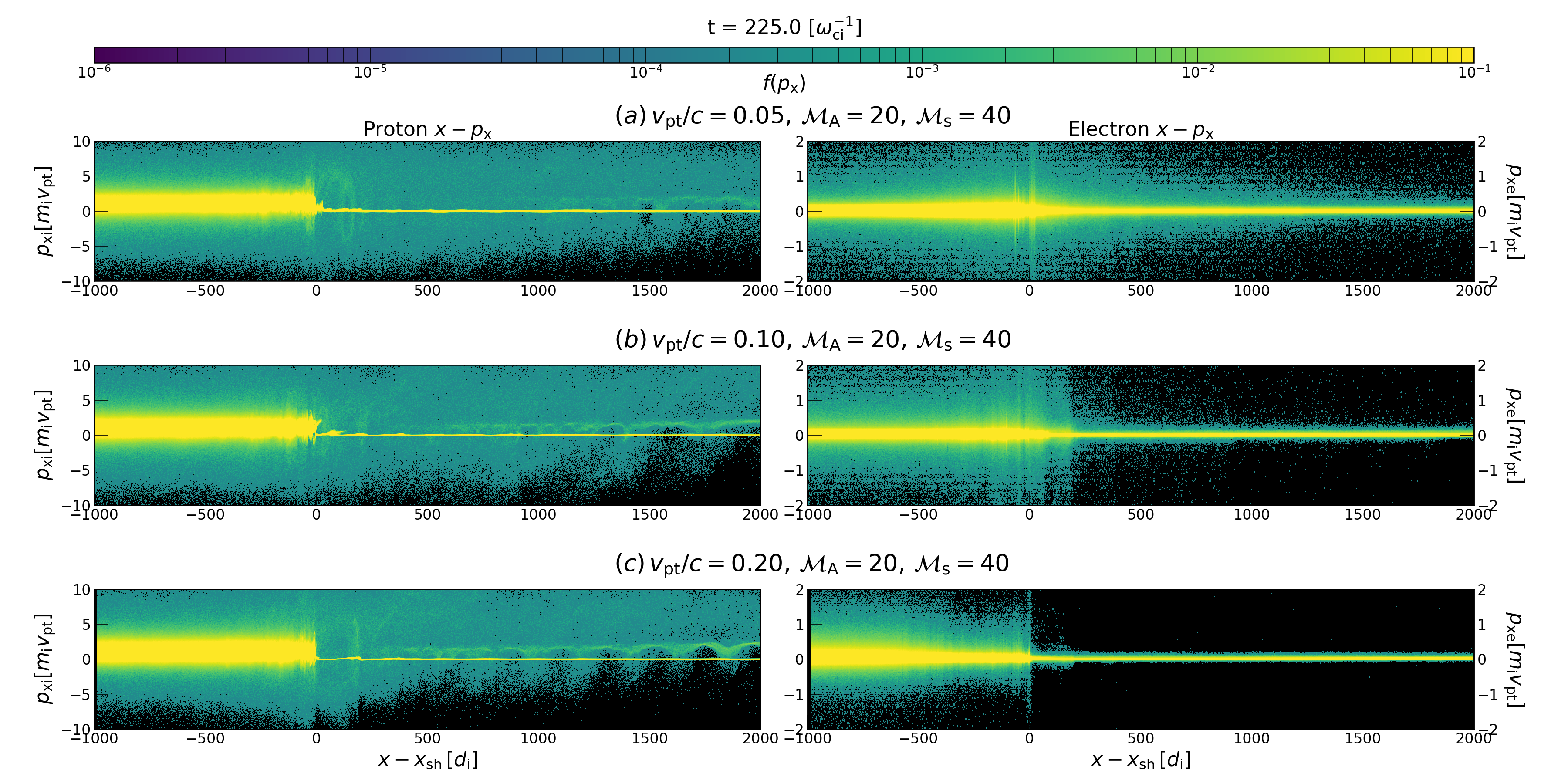}
\caption{
Snapshot of proton and electron $x-p_{\rm x}$ phase spaces at $t=225\, \omega_{\rm ci}^{-1}$ for $v_{\rm pt}/c=0.05,0.1,$ and $0.2$ (top to bottom), where other parameters, $\mathcal{M}_{\rm A}=20,\mathcal{M}_{\rm s}=40,$ and $m_{\rm R}=100$, are fixed. 
While the proton phase spaces (left panels) are quite similar for different $v_{\rm pt}$, there are fewer upstream electrons (right panels) for larger shock speeds.}
\label{fig:x-vx-vp}
\end{figure*}

In all of the panels of Figure \ref{fig:magtheta}, left and right axes indicate the magnetic field (red) and the effective $\theta_{\rm Bn}$ (blue), respectively,
where the thickness of the lines corresponds to different  runs;
since the precursors exhibit rapid fluctuations, we present profiles averaged in time over $100-150\,\omega_{\rm ci}^{-1}$. 
As the particles cannot outrun the shock if the inclination exceeds a critical angle $\theta_{\rm *}\equiv \cos^{-1}(v_{\rm sh}/c)$, we also display $\theta_{\rm *}$.
As the wavelength of dominant EM modes in resonant/nonresonant streaming instability scales with $v_{\rm d}/v_{\rm A}$ (where $v_{\rm d}\sim v_{\rm sh}$ is the drift speed of the streaming particles), the horizontal axis showing the distance from the shock is normalized to $\mathcal{M}_{\rm A} d_{\rm i}$.

Figure \ref{fig:magtheta}(a) shows three runs that differ by the piston speed: $v_{\rm pt}/c=0.05, 0.1,$ and $0.2$, respectively, with $\mathcal{M}_{\rm A}=20, \mathcal{M}_{\rm s}=40$, and $m_{\rm R}=100$ fixed (the runs $\mathcal{A}1$, $\mathcal{B}1$, and $\mathcal{C}1$ in Table \ref{tab:simpara}).
The profiles of the magnetic field and $\theta_{\rm Bn}$ are almost identical because magnetic field amplification mainly depends on the Alfv\'en Mach number and the proton acceleration efficiency, such dependences are explained as follows.

Recent kinetic simulations have shown that the ratio of the anisotropic momentum flux of driving particles ($\mathcal{P}_{\rm nt}$) to the initial magnetic pressure ($P_{\rm B,0}$) regulates the saturation of the nonresonant streaming instability \citep{gupta+21,zacharegkas+24}.
Since in the upstream frame $\mathcal{P}_{\rm nt} \approx n_{\rm nt} \langle p_{\rm nt,x}\rangle v_{\rm d}$ and $P_{\rm B,0} = 0.5 \rho v_{\rm A}^2$, the final $\delta B/B_{\rm 0}$ is given as
\begin{equation}\label{eq:Bampli}
    \frac{\delta B}{B_{\rm 0}} \approx \sqrt{\frac{1}{2} \frac{n_{\rm nt}}{n_{\rm 0}}\frac{\langle p_{\rm nt,x}\rangle}{m_{\rm i}v_{\rm pt}}}\,\mathcal{M}_{\rm A}\,
\end{equation}
where $\langle p_{\rm nt,x}\rangle$ is the average $x$-momentum of NT protons in the upstream rest frame and $n_{\rm nt}/n_{\rm 0}$ represents the ratio of NT protons to thermal protons in the upstream.
Here we have assumed the bulk speed of NT protons $v_{\rm d}\sim v_{\rm sh}$, as justified in Appendix \ref{app:Bfieldmi} (see also Appendix \ref{app:resonantvsnonresonant}). 
Clearly, when $\mathcal{M}_{\rm A}$ is fixed (as in Figure \ref{fig:magtheta}(a)), the profiles of $B_{\rm \perp}/B_{\rm 0}$ are not expected to depend much on $v_{\rm pt}$.

To understand how the results depend on the shock Mach numbers $\mathcal{M}_{\rm A}$ and $\mathcal{M}_{\rm s}$, and on the plasma $\beta$, we consider Figures \ref{fig:magtheta}(b)-(d), where $v_{\rm pt}/c=0.1$ and $m_{\rm R}=100$ are fixed.
In general, we find that the magnetic field amplification drops for low Mach number shocks, as expected from Equation \ref{eq:Bampli}.
However, the dependence of $B_{\rm \perp}/B_{\rm 0}$ on $\mathcal{M}_{\rm s}$ for a fixed $\mathcal{M}_{\rm A}$, shown in Figure \ref{fig:magtheta}(c), is non-trivial and suggests that the magnetic field at the shock is sensitive to the temperature of the upstream plasma, an effect that may affect the injection efficiency as well as the evolution of the spectra, as we discuss in \S \ref{sec:accsig}.
\begin{figure}
    \centering
    \includegraphics[width=2.9in]{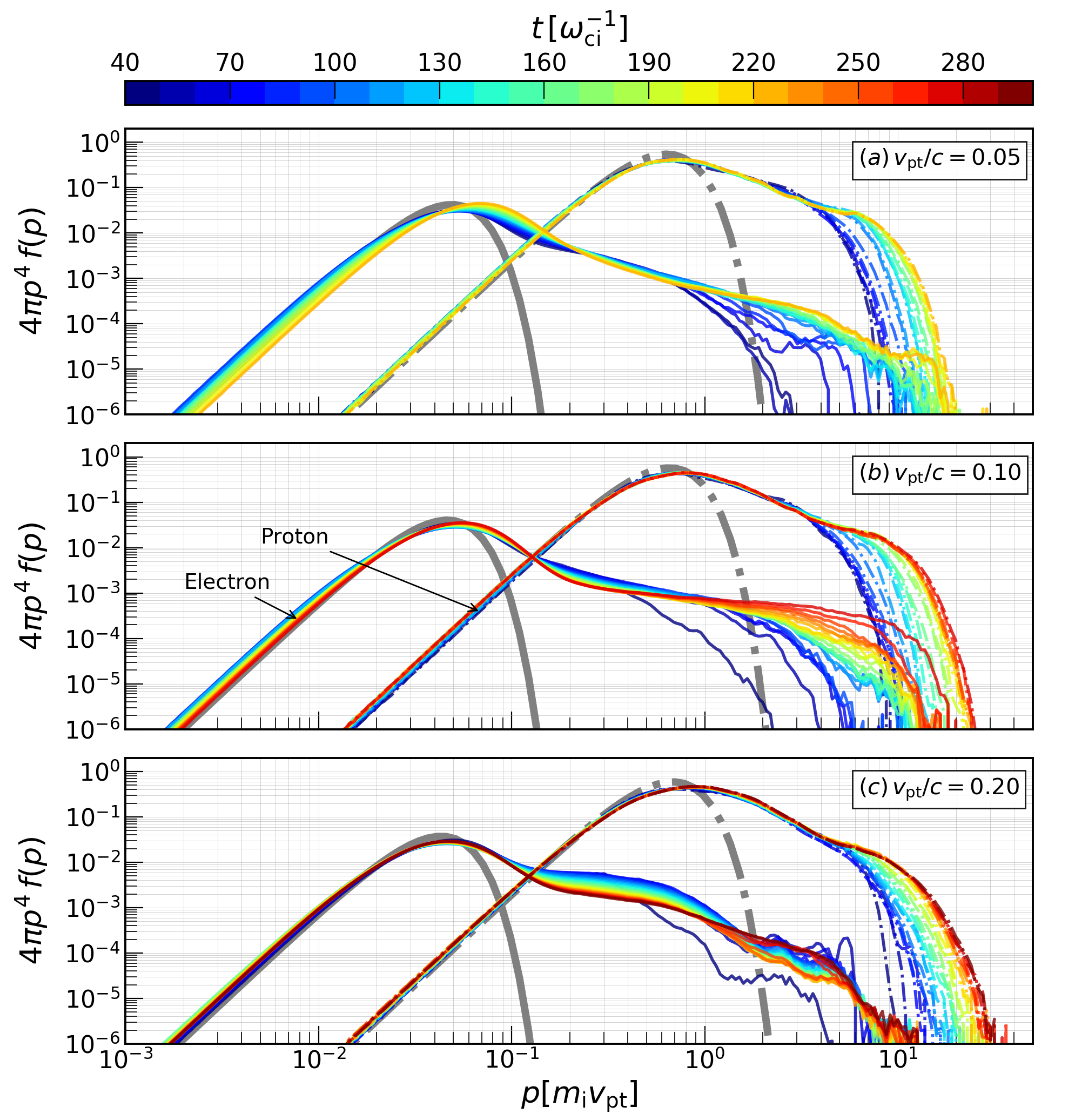}
    \caption{
    Time evolution of downstream proton (dashed-dotted lines) and electron (solid lines) spectra for $v_{\rm pt}/c=0.05,0.1,$ and $0.2$,
    The designations of different panels are kept identical to Figure \ref{fig:x-vx-vp}.
    The grey curves show the fitted thermal Maxwellian.
    Both protons and electrons show the NT tail.
    Panel (c) shows that the electron NT tail stalled.}
    \label{fig:spec_vp}
\end{figure}

In Figures \ref{fig:magtheta}(e) and (f), we investigate the impact of the reduced mass ratio for either low $\mathcal{M}_{\rm A}=5$ (Figure \ref{fig:magtheta}(e), runs $\mathcal{B}2$ and $\mathcal{B}2_{\rm II}$) or high $\mathcal{M}_{\rm A}=20$ (Figure \ref{fig:magtheta}(f), runs $\mathcal{B}1$, $\mathcal{B}1_{\rm II}$, $\mathcal{B}1_{\rm III}$, and $\mathcal{B}1_{\rm IV}$);
the shock speed $v_{\rm pt}/c$ and the proton sonic Mach number $\mathcal{M}_{\rm s}$ are fixed to $0.1$ and $40$ respectively.
Both panels show that the profiles for different mass ratios are similar, though a moderate increase in amplification is noticed for small $m_{\rm R}$ runs, not predicted by Equation \ref{eq:Bampli}.
Upon a thorough investigation, we have identified that at $t=100-150\omega_{\rm ci}^{-1}$ the nonresonant streaming instability has not yet saturated for the larger $m_{\rm R}$.
Since for a fixed $\mathcal{M}_{\rm A}$ the magnetization of electrons increases with $m_{\rm R}$, self-generated perturbations evolve more slowly and the streaming instability is slower, as we have confirmed via the controlled runs shown in Appendix \ref{app:Bfieldmi}.

%
\section{Acceleration Signatures}\label{sec:accsig}
\begin{figure*}[ht!]
\centering
\begin{minipage}{.32\linewidth}
\centering
\includegraphics[width=2.25in]{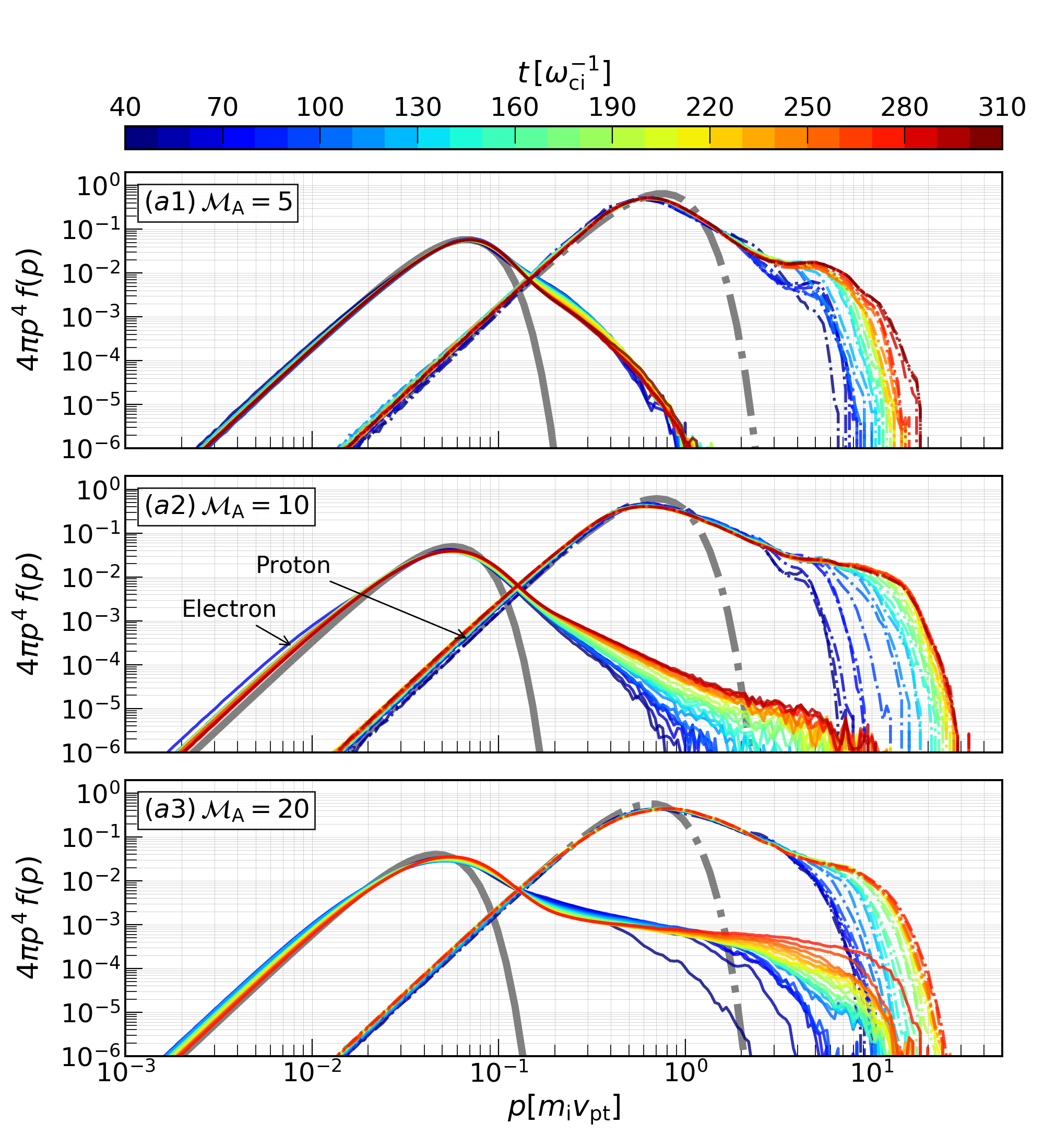}
\end{minipage}
\begin{minipage}{.32\linewidth}
\centering
\includegraphics[width=2.25in]{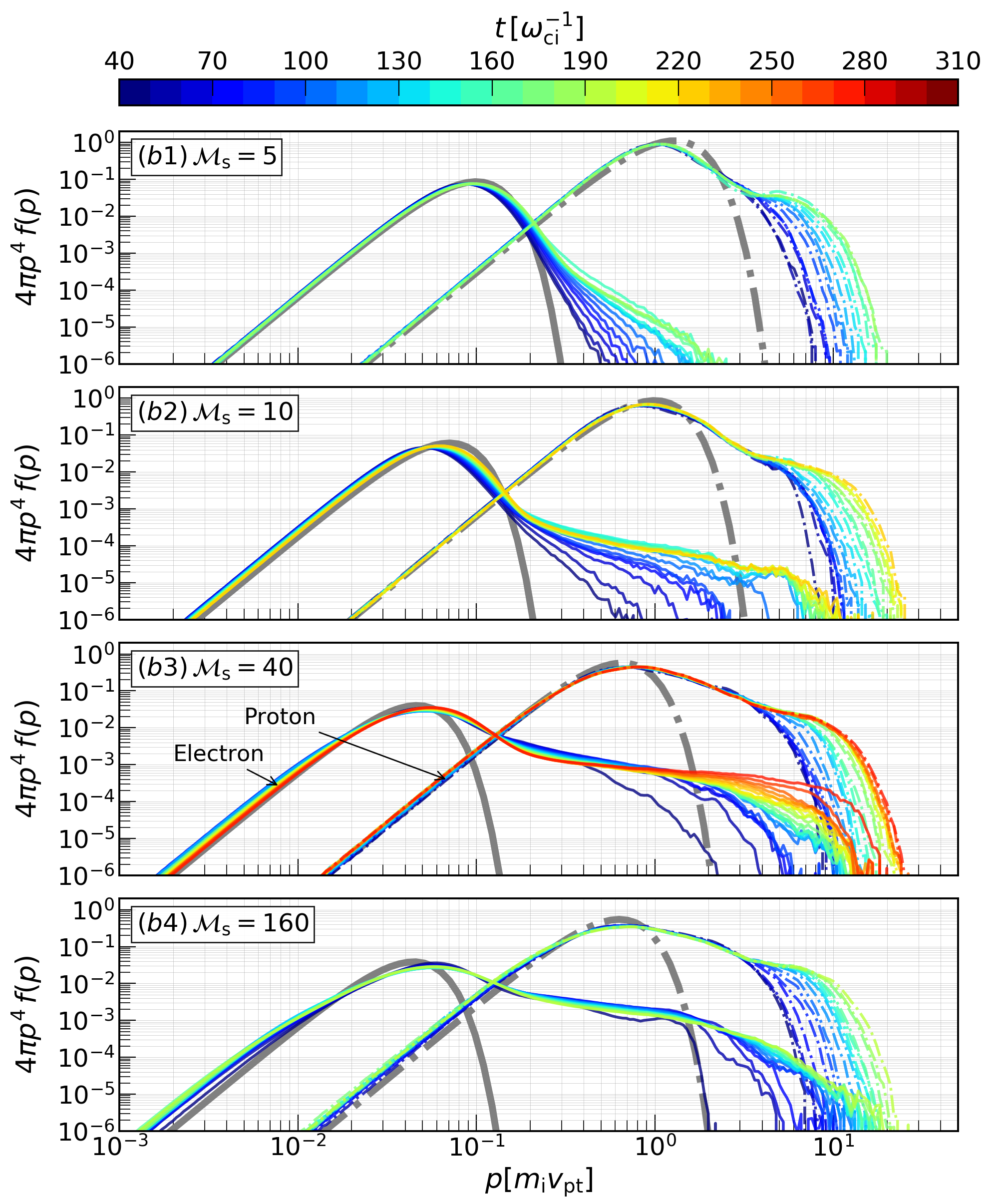}
\end{minipage}
\begin{minipage}{.32\linewidth}
\centering
\includegraphics[width=2.25in]{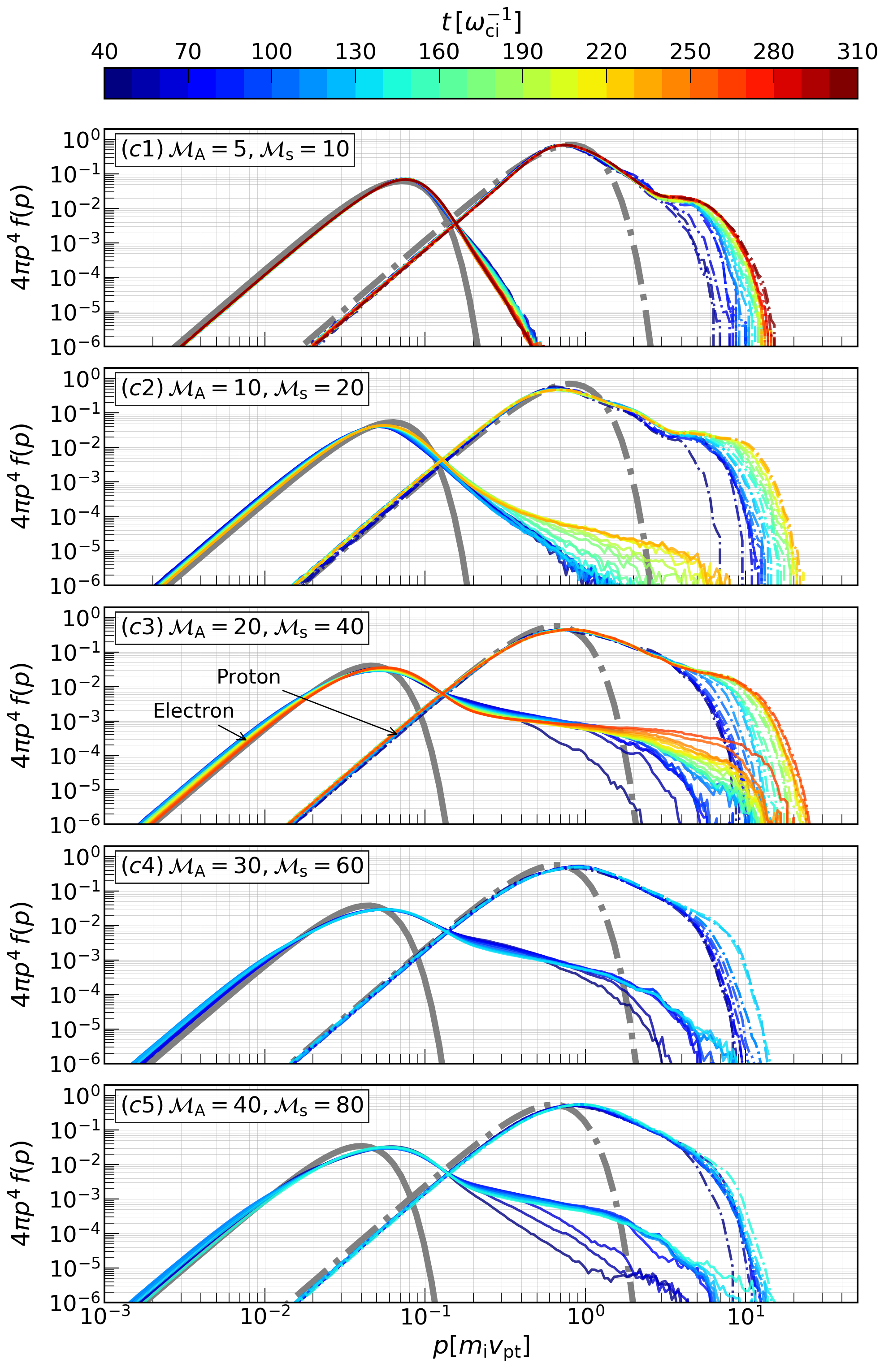}
\end{minipage}
\caption{
Downstream spectra for different $\mathcal{M}_{\rm A}$ while keeping $\mathcal{M}_{\rm s}=40$ fixed (left panels), a fixed $\mathcal{M}_{\rm A}=20$ for different $\mathcal{M}_{\rm s}$ (middle panels), and different $\mathcal{M}_{\rm A}=\mathcal{M}_{\rm s}/2$ but a fixed $\beta=1$ (right panels).
For all panels, $v_{\rm pt}=0.1\,c$ and $m_{\rm R}=100$.
{\it Left panels}: Electron NT tail differs remarkably for different $\mathcal{M}_{\rm A}$. 
Besides the run with $\mathcal{M}_{\rm A}=5$ (panel a1), the spectra are evolving to higher energies (i.e., momentum $p\gtrsim m_{\rm i}v_{\rm pt}$).
{\it Middle panels}: the normalization of NT tail is sensitive to $\mathcal{M}_{\rm s}$.
{\it Right panels}: evolution is similar to panels (a2) and (a3), though at a slower rate compared to the same $\mathcal{M}_{\rm A}$ runs.}
\label{fig:spec-Ma-Ms-MaMs}
\end{figure*}

\begin{figure*}[ht!]
\centering
\tabskip=0pt
\valign{#\cr
  \hbox{%
    \begin{minipage}[b]{.45\textwidth}
    \centering
    \includegraphics[width=3.1in]{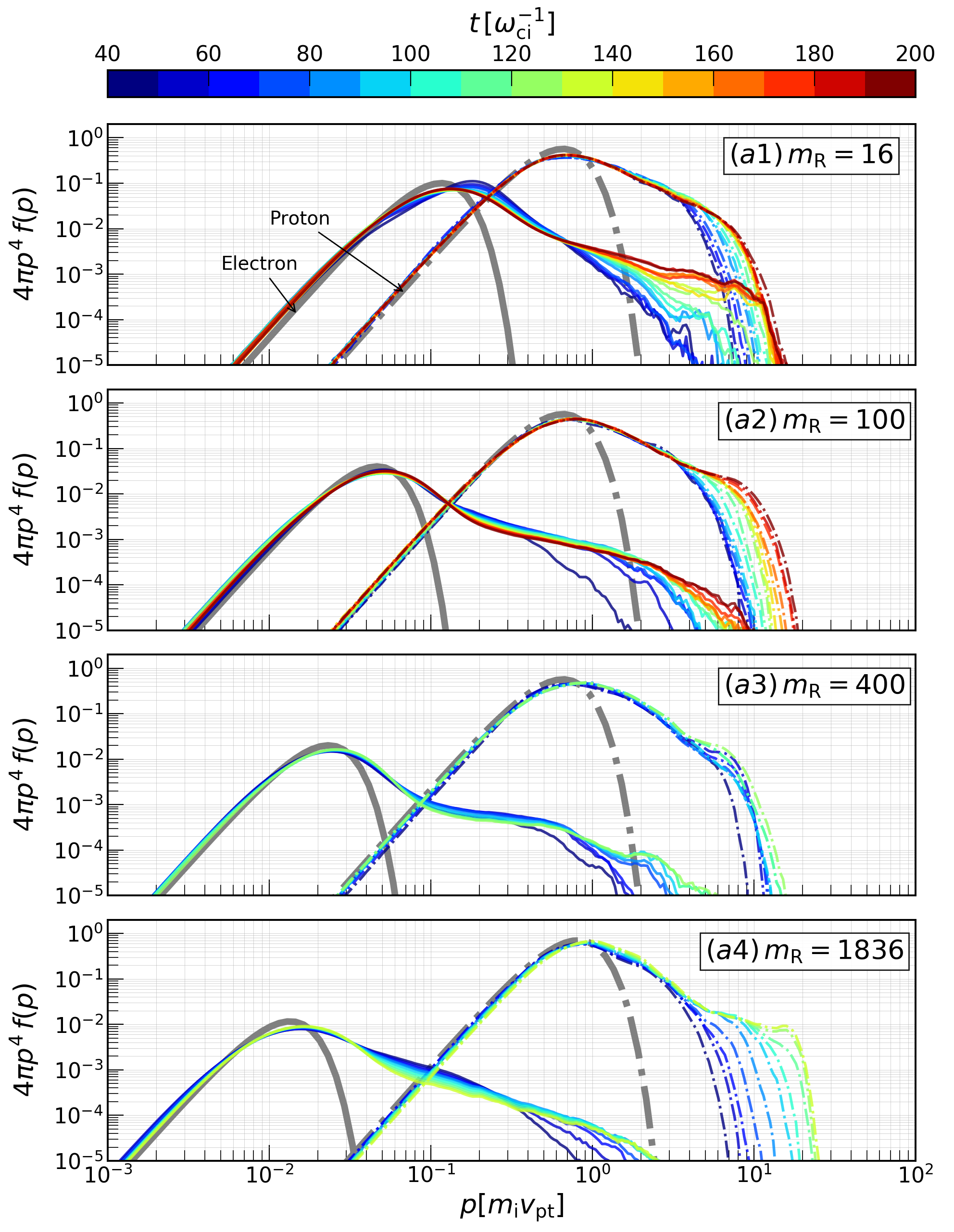}
    \end{minipage}%
  }\cr
  \noalign{\hfill}
  \vspace{.125\textwidth}
  \hbox{%
    \begin{minipage}{.45\textwidth}
    \centering
     \includegraphics[width=3.1in]{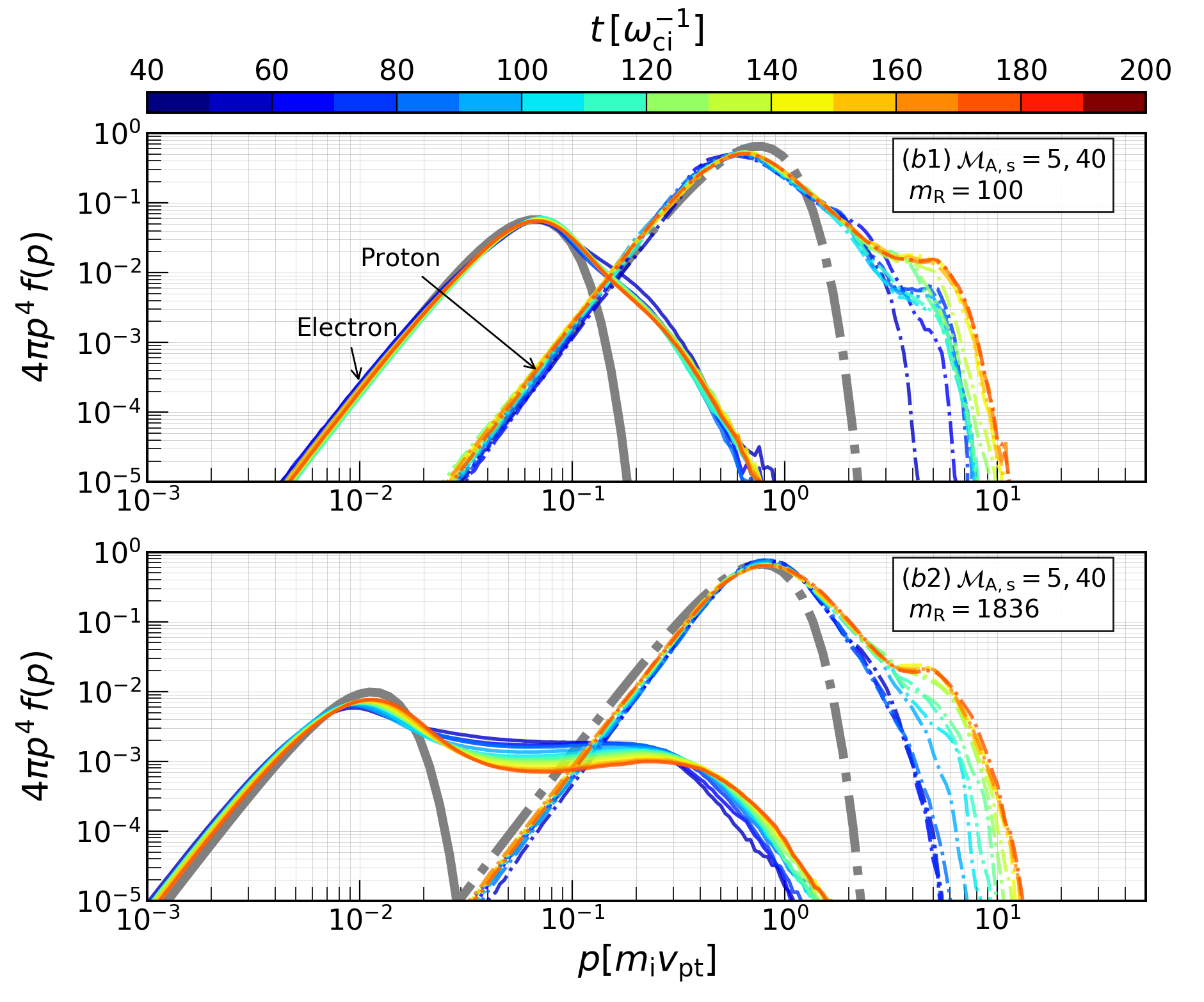}
    \end{minipage}%
  }\vfill
  \hbox{%
    \begin{minipage}{.45\textwidth}
    \centering
    \end{minipage}%
  }\cr
}
\caption{Time evolution of downstream spectra for simulations with different mass ratios, $m_{\rm R}$.
For panels (a1)--(a4): $v_{\rm pt}/c=0.1,\mathcal{M}_{\rm A}= \mathcal{M}_{\rm s}/2=20$ ($\mathcal{B}1$ series in Table \ref{tab:simpara}), for panels (b1)--(b2): $v_{\rm pt}/c=0.1,\mathcal{M}_{\rm A}=\mathcal{M}_{\rm s}/8=5$ ($\mathcal{B}2$ series in Table \ref{tab:simpara}).
As $m_{\rm R}$ increases, the peak of the electron thermal distribution moves to smaller momenta, as expected (\S \ref{subsec:spec}).
}
\label{fig:spec-mr}
\end{figure*}

To investigate particle acceleration, we consider the $x-p_{\rm x}$ phase space of electrons and protons.
Since simulations are performed in the upstream frame, positive (negative) $p_{\rm x}$ populations ahead of the shock correspond to particles escaping from (returning to) the shock, a signature of DSA.
As in the previous section, we first show results for different piston/shock speeds and then discuss their dependence on Mach numbers and mass ratio in Figures \ref{fig:x-vx-vp}-\ref{fig:spec-mr}.

\subsection{Shock Speed}\label{subsec:vp}
Figure \ref{fig:x-vx-vp} shows the $x-p_{\rm x}$ phase-space  at $t=225\,\omega_{\rm ci}^{-1}$ for three runs with $v_{\rm pt}/c=0.05,0.1$, and $0.2$.
Left panels, representing the proton phase-space, do not show significant differences, except that the upstream free-escape boundary, loosely defined as a location beyond which the energetic populations are no longer confined by magnetic fluctuations and thus unable to return to shock, shrinks with increasing shock speed;
in other words, the extent of the precursor gets shorter in high-speed shocks.

The differences in the phase-space are more prominent for electrons, shown in the right panels of Figure \ref{fig:x-vx-vp}.
Clearly, there are fewer energetic electrons for high shock speed, and they are hardly noticeable for $v_{\rm pt}/c=0.2$, consistent with the findings of the previous section: for larger shock speeds, shocks tend to become superluminal frequently, which allows particles to outrun the shock only for a smaller range of pitch angles (see the blue thick lines in Figure \ref{fig:magtheta}(a)).
1D simulations tend to exaggerate this effect because they do not allow for a change in $B_x$ and for shock rippling, and in fact, 2D simulations of trans-relativistic shocks do accelerate particles both protons and electrons \citep[e.g.,][]{crumley+19}. 
Future investigation is necessary to validate whether the trend of precursor shortening with increasing shock speeds persists in multidimensions.

The features mentioned above are also observable in the downstream spectra, measured in a region of $1000\,d_{\rm i}$ behind the shock, as shown 
in Figure \ref{fig:spec_vp}.
While the evolution of electron spectra is qualitatively similar for $v_{\rm pt}/c=0.05$ and $0.1$ shocks, the downstream electron spectrum for $v_{\rm pt}/c=0.2$ develops faster, but its NT tail saturates at $p \simeq 10\,m_{\rm i}v_{\rm pt}$ after the shock becomes superluminal.
\subsection{Mach Number}\label{subsec:M}
Figure \ref{fig:spec-Ma-Ms-MaMs} displays the evolution of spectra of electrons and protons for three different sets of runs, which differ by $\mathcal{M}_{\rm A}$ (Figures \ref{fig:spec-Ma-Ms-MaMs}(a1)--(a3)), $\mathcal{M}_{\rm s}$ (Figures \ref{fig:spec-Ma-Ms-MaMs}(b1)--(b4)), and both $\mathcal{M}_{\rm A}$ and $\mathcal{M}_{\rm s}/2$ (Figures \ref{fig:spec-Ma-Ms-MaMs}(c1)--(c5));
note that these runs span different values of $\beta\equiv 4(\mathcal{M}_{\rm A}/\mathcal{M}_{\rm s})$.
For all these runs we have used a fixed $v_{\rm pt}/c=0.1$ and $m_{\rm R}=100$.
In general, phase-space distributions do not differ much from those in Figure \ref{fig:x-vx-vp}, except for the energy of reflected electrons, as outlined below.

In low-Mach number shocks, the upstream $x-p_{\rm x}$ phase-space is mostly populated by electrons with $p_{\rm x}>0$, i.e., shock-reflected electrons struggle to return to the shock;
also, the extent of their momentum distribution is smaller than that found in high Mach shocks.
This is due to the fact that a weaker self-generated turbulence with $\delta B/B\lesssim 1$ makes the scattering less effective and leads to a smaller fractional energy gain.

Therefore, electron acceleration may stall either because the shock becomes superluminal (at high speeds, see Figure \ref{fig:x-vx-vp}), or because of the lack of self-generated turbulence and inefficient pre-acceleration mechanisms (for low Mach numbers). 
This is better seen from the downstream spectra, as we discuss in the following.

All panels in Figure \ref{fig:spec-Ma-Ms-MaMs} show a distinct power-law tail in momentum for both electrons and protons.
However, unlike for protons, the slope of the electron spectrum can vary dramatically: 
only in shocks with $\mathcal{M}_{\rm A,s}\gtrsim 10$ such a slope matches the DSA prediction $\sim 4$, with the tail evolving to larger and larger energies with time.
We note that the spectral evolution is similar for a given $\mathcal{M}_{\rm A}$, but acceleration is less rapid for lower $\mathcal{M}_{\rm s}$ shocks;  
see, e.g., Figures \ref{fig:spec-Ma-Ms-MaMs}(b2) and (b3), which have the same $\mathcal{M}_{\rm A}=20$, but $\mathcal{M}_{\rm s}=10$ and $40$, respectively.
As long as the plasma is not too hot (i.e., $\beta$ is not much larger than one), the amplitude of the upstream $\delta B/B$ that controls particle confinement is regulated mostly by $\mathcal{M}_{\rm A}$ and depends weakly on $\mathcal{M}_{\rm s}$,
causing a similar evolution of the NT tail among these runs.
Interestingly, unlike for low $\mathcal{M}_{\rm A}$ shocks, the maximum momentum keeps growing with time.

We conclude that low $\mathcal{M}_{\rm s}$ shocks are capable of injecting thermal electrons into DSA and of accelerating them to higher energies as long as $\mathcal{M}_{\rm A}$ is large enough.
However, if \(\mathcal{M}_{\rm s} \ll 10\), both the efficiency of particle acceleration and the development of self-generated magnetic fields may be affected due to the lack of confinement of NT particles across the shock.
This has interesting implications for understanding the NT $X$-ray and radio emissions from galaxy clusters, which typically feature shocks with $\mathcal{M}_{\rm s}\lesssim 5$ in high-$\beta$ plasmas \citep[e.g.,][]{brunetti+14, vanweeren+17}.  
\subsection{Mass Ratio}\label{subsec:mR}
This section deals with assessing the role of the reduced mass ratio $m_{\rm R}$ in regulating electron DSA.
Since the computational cost increases drastically with $m_{\rm R}$, we ran selected simulations for two different values of $\mathcal{M}_{\rm A}$ while keeping $v_{\rm pt}/c=0.1$ and $\mathcal{M}_{\rm s}=40$ fixed.

\begin{figure*}[ht!]
    \centering
   \includegraphics[width=5.5in]{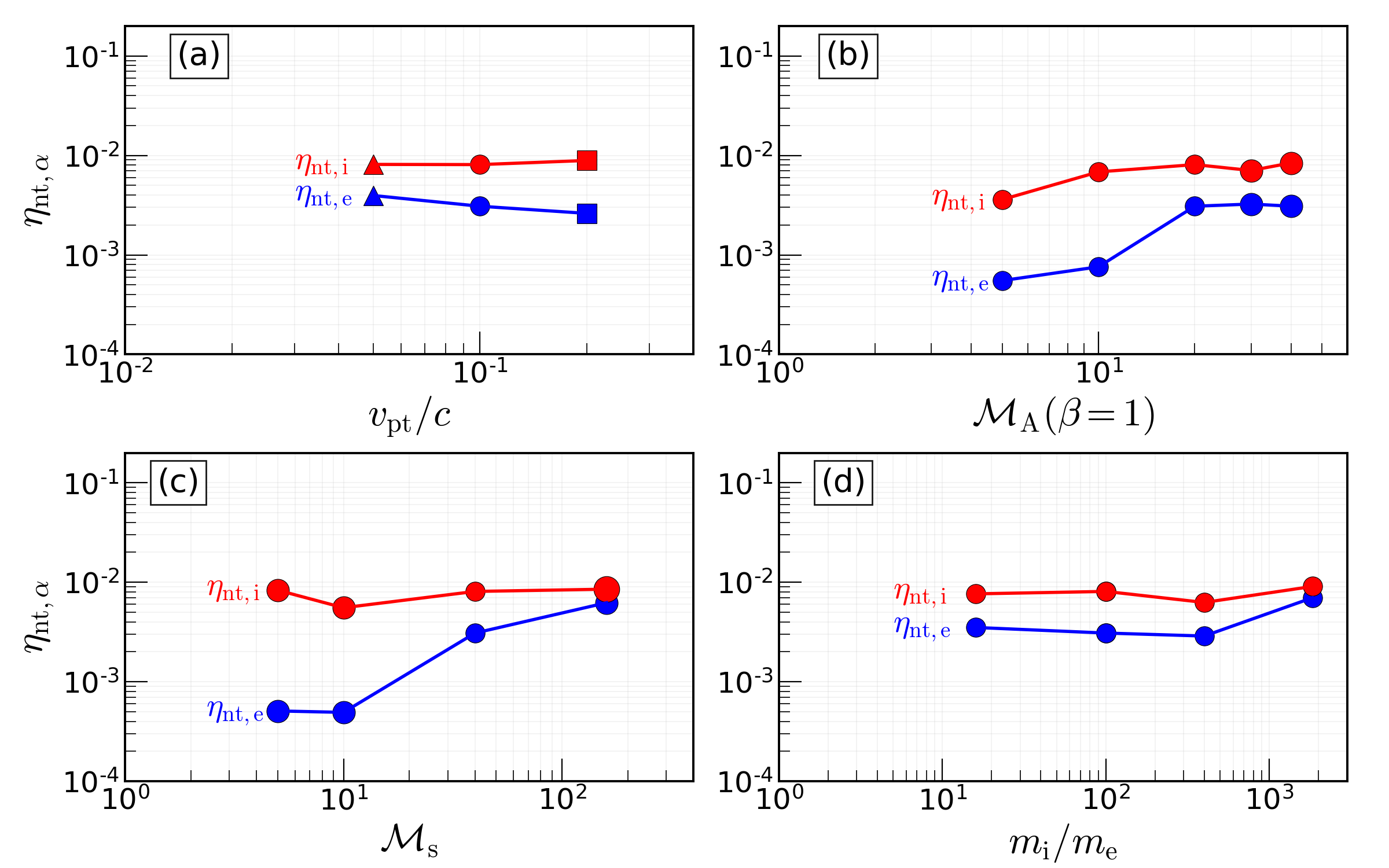}
   \caption{
Injection efficiency of energetic particles as a function of $v_{\rm pt}\equiv 3v_{\rm sh}/4$ (panel a), $\mathcal{M}_{\rm A}$ (panel b), $\mathcal{M}_{\rm s}$ (panel c), and $m_{\rm i}/m_{\rm e}$ (panel d).
Red and blue lines represent the fraction of energetic particles with momentum $p\geq p_{\rm 0,\alpha}$ (Equation \ref{eq:p0alpha}), calculated at late times over a region of $1000\,d_{\rm i}$ immediately downstream of the shock.
Unless otherwise specified, $v_{\rm pt}=0.1\,c$, $\mathcal{M}_{\rm A}=20$, $\mathcal{M}_{\rm s}=40$, and $m_{\rm R}=100$.
}\label{fig:eta_energetic}
\end{figure*}
Let us first focus on runs with four different mass ratios and a fixed $\mathcal{M}_{\rm A}=20$, as displayed in Figures \ref{fig:spec-mr}(a1)--(a4). 
Comparison of the electron spectra among these runs suggests that for smaller $m_{\rm R}$ the downstream thermal electrons are more energetic, as expected since the thermal electron distribution measured in unit of $p/m_{\rm i}v_{\rm pt}$ scales as $\sim 1/\sqrt{m_{\rm R}}$ (see \S\ref{subsec:spec}). 
All four cases show the development of NT electrons, implying that altering the mass ratio does not halt acceleration for large $\mathcal{M}_{\rm A}$; indeed, both NT electrons and protons continue to increase energy.

Instead, for a lower $\mathcal{M}_{\rm A}=5$ with all other parameters identical to Figures \ref{fig:spec-mr}(a1) and (a4), the shape of the NT spectra are quite dependent on the mass ratio, as shown in Figures \ref{fig:spec-mr}(b1) and (b2).
A comparison between Figures \ref{fig:spec-mr}(b1) and (b2) reveals that thermal electrons appear more energetic in the $m_{\rm R}=100$ case compared to the $m_{\rm R}=1836$ case, as expected.
While for the large $m_{\rm R}$ run the NT electron spectrum is consistent with the DSA prediction, the NT tail appears steeper in the small $m_{\rm R}$ case.
For low $\mathcal{M}_{\rm A}$ shocks, the downstream thermal electron distribution for small mass ratio runs is not far from the cut-off of the maximum energy, causing the NT spectrum to appear steeper than that found in large $m_{\rm R}$ runs.
%
\section{Physical Scalings}\label{sec:scalings}
In this section we discuss how our simulations can inform the electron DSA parameters used in modeling the NT phenomenology of several galactic and extragalactic sources \citep[see, e.g.,][]{morlino+12,gupta+18,winner+20,reynolds+21}, in particular the fraction of electrons injected into DSA (\S \ref{subsec:eta}) and the electron acceleration efficiency (\S \ref{subsec:epsilon}).
Since modeling shock-powered systems often involve knowing the magnetic energy density and the postshock electron temperature (e.g., to calculate thermal X-ray and synchrotron radio/X-ray emission), we have estimated these quantities as well (\S \ref{subsec:etheB} and \S\ref{subsec:TeTp}).
Results are presented as a function of different shock parameters: 
we fix $v_{\rm pt}=0.1\,c$, $\mathcal{M}_{\rm A}=\mathcal{M}_{\rm s}/2=20$, and $m_{\rm R}=100$ as default parameters, varying $v_{\rm pt}$, $\mathcal{M}_{\rm A}$, $\mathcal{M}_{\rm s}$, and $m_{\rm R}$ in the four panels (a)--(d) of Figures \ref{fig:eta_energetic}--\ref{fig:tempds}, respectively.
\subsection{Injection Efficiency and Electron-to-Proton Ratio} \label{subsec:eta}
%
One of the main goals of an acceleration theory is to determine the fraction of particles that become NT.
We therefore introduce the following expression:
\begin{equation}
    \eta_{\rm nt, \alpha}=\frac{\int_{p_{\rm 0,\alpha}}dp\, 4\pi  p^2\, f_{\rm \alpha}(p)}{\int dp\, 4\pi  p^2\, f_{\rm \alpha}(p)}  
\end{equation}
where $f_{\rm \alpha}(p)$ is the distribution function of thermal + NT particles in the downstream, $p_{\rm 0,\alpha}$ represents the boundary between thermal and NT particles for the species $\alpha=e,i$ (Equation \ref{eq:p0alpha}).

Figure \ref{fig:eta_energetic} shows $\eta_{\rm nt, i,e}$ in red and blue, respectively.
The proton injection efficiency $\eta_{\rm nt, i}\simeq 10^{-2}$ does not show a significant dependence on the shock speed (Figure \ref{fig:eta_energetic}(a)) and mass ratios $m_{\rm R}$ (Figure \ref{fig:eta_energetic}(d));
only for low $\mathcal{M}_{\rm A}$ or $\mathcal{M}_{\rm s}$, $\eta_{\rm nt, i}$ decreases by a factor of $\lesssim 2$ (Figures \ref{fig:eta_energetic}(b) and (c)).

The electron injection efficiency (blue lines in Figure \ref{fig:eta_energetic}) is $\eta_{\rm nt, e}\lesssim 0.5\times 10^{-2}$, comparable to protons', and does not show much dependence on shock speed and $m_{\rm R}$,
as shown in Figures \ref{fig:eta_energetic}(a) and (d), respectively. 
However, for low Mach number shocks $\eta_{\rm nt, e}$ drops by a factor of $\sim 5$ (Figures \ref{fig:eta_energetic}(b) and (c)), i.e., the electron acceleration is more sensitive to the shock Mach number than protons'.
Thus, we conclude that high-Mach number quasi-parallel shocks ($\mathcal{M}_{\rm s},\mathcal{M}_{\rm A}\gtrsim 10$) are generally efficient in injecting protons and electrons into DSA.

Another crucial parameter that we can estimate from the above analysis is the electron-to-proton number ratio, $K_{\rm ei}$, measured at a fixed NT energy.
As the NT tail of electron and proton spectra starts at two different momenta, $p_{\rm 0,i}$ and $p_{\rm 0,e}$, the ratio at a given momentum $p$ would be
\begin{equation}\label{eq:kei}
K_{\rm ei} \equiv \frac{f_{\rm e}(p)}{f_{\rm i}(p)}\approx \frac{\eta_{\rm nt, e}}{\eta_{\rm nt, i}}\,\left(\frac{p}{p_{\rm 0,i}}\right)^{q_{\rm i}-q_{\rm e}}\,m_{\rm R}^{-(q_{\rm e}-3)/2}\,
\end{equation}
where $q_{\rm e,i}$ is the slope of the NT electron/proton distributions, and $p_{\rm 0,e}\approx p_{\rm 0,i}/\sqrt{m_{\rm R}}$ as in Equation \ref{eq:p0alpha}.
Equation \ref{eq:kei} suggests that in PIC simulations $K_{\rm ei}$ is expected to drop with increasing $m_{\rm R}$. 
For $m_{\rm R}=1836$, assuming $q_{\rm e,i}\approx 4$, we find $K_{\rm ei}\lesssim 10^{-2}$.
This is expected since electrons have to go through a larger number of Fermi cycles before they can achieve a momentum $p$, which is normally taken larger than $p_{\rm 0,i}$.

To summarize, at quasi-parallel shocks, the injection efficiency of NT particles ($p\geq p_{\rm 0,\alpha}$), both protons and electrons, is $\eta_{\rm nt, i,e}\approx 10^{-2}$, with $\eta_{\rm nt, e}$ only marginally smaller than $\eta_{\rm nt, i}$.
Since the proton injection fraction $\eta_{\rm nt, i}\sim 10^{-2}$, the normalization of electron-to-proton DSA tail at a given NT momentum is expected to be $K_{\rm ei}\lesssim 10^{-2}$, which is consistent with observations of supernova remnants \citep[e.g.,][]{berezhko+04a, morlino+12, slane+14} and in general agreement with the electron/proton ratio measured in Galactic CRs.
Moreover, we find that the ratio $K_{\rm ei}$ becomes smaller in lower Mach number shocks.
\subsection{Energy budgets}
\label{subsec:epsilon}
%
Next, we consider the energy density in different populations using the traditional definition
\begin{eqnarray}\label{eq:effengy}
  \epsilon_{\rm \alpha}\equiv\frac{1}{e_{\rm tot}}\int dp \, 4\pi  p^2\, f_{\alpha}(p) \,(\gamma-1)m_{\rm \alpha}c^2
\end{eqnarray}
where, $e_{\rm tot}$ is the sum of the internal energy density in the thermal and NT populations ($\sum_{\rm \alpha} \int_{0}^{\infty}\,dp\,4\pi  p^2\, f_{\alpha}(p)\,(\gamma-1)m_{\rm \alpha}c^2$) and the magnetic energy density ($e_{\rm B}=B^2/8\pi$) in the downstream.
Thus, $\epsilon_\alpha$ represents the fraction of the total postshock energy density available to different populations.

\begin{figure*}
    \centering
    \includegraphics[width=5.5in]{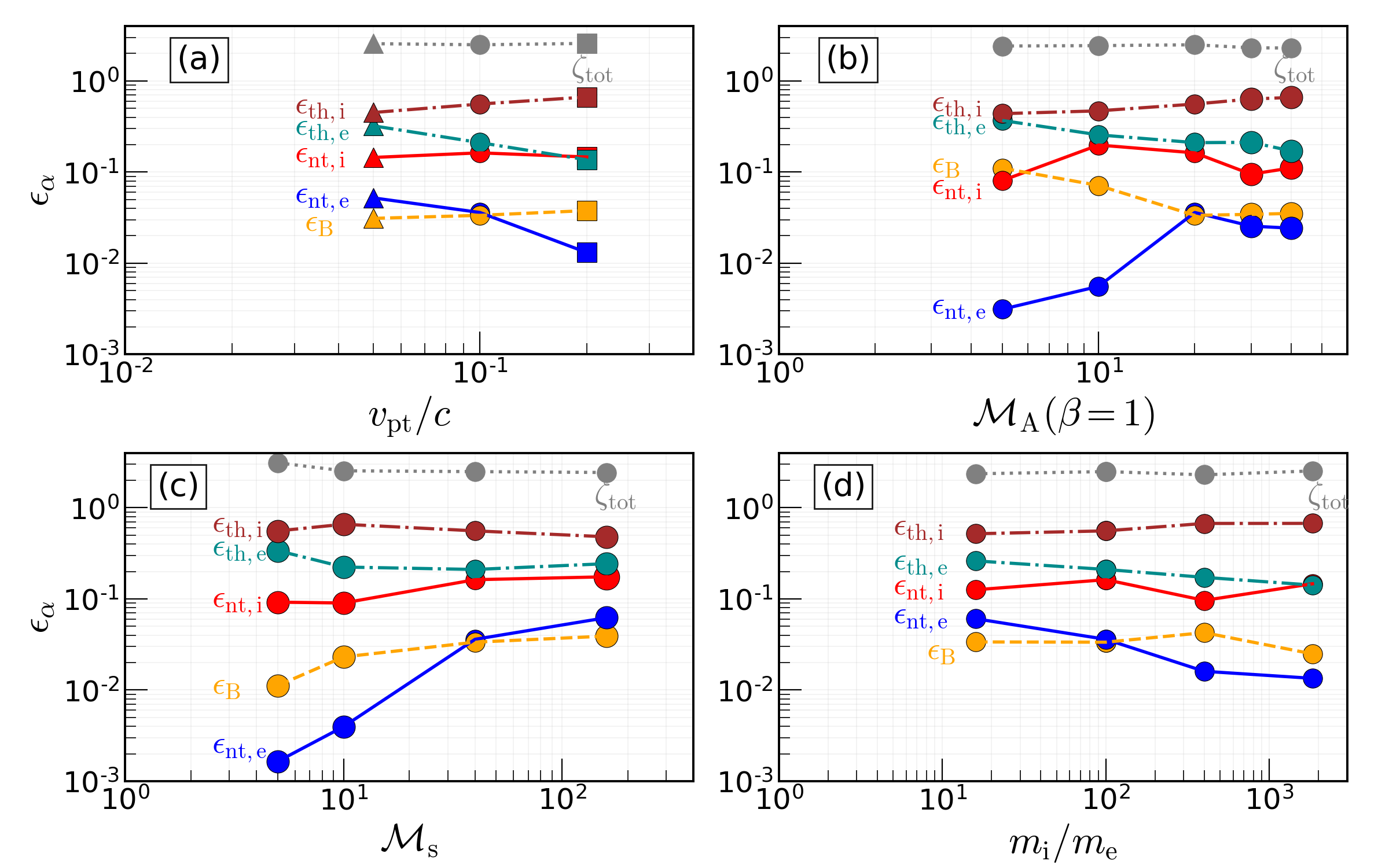}
\caption{
Energy budget for the downstream thermal and nonthermal populations, and the magnetic fields. 
The red and blue curves represent the fraction of total postshock energy density in NT protons and electrons with momentum $p\geq p_{\rm 0,\alpha}$ (as described by Equations \ref{eq:p0alpha} and \ref{eq:effengy}). 
$\epsilon_{\rm th, i}$ and $\epsilon_{\rm th, e}$ represent the energy density fractions for thermal protons (brown) and electrons (cyan), obtained for $p<p_{\rm 0,\alpha}$, and $\epsilon_{\rm B}$ is the magnetic energy density.
}
\label{fig:epsilon}
\end{figure*}

In Figure \ref{fig:epsilon}, the grey curves show the ratio of $e_{\rm tot}$ to the bulk kinetic energy density of the upstream flow, $\zeta_{\rm tot}\equiv e_{\rm tot}/(0.5\rho_1 v^2_{\rm sh})$, in our simulations. 
Note that $\zeta_{\rm tot}>1$ does not violate energy conservation, since the downstream total energy density per particle, $\zeta_{\rm tot}/\mathcal{R}$, is less than unity (where $\mathcal{R}\approx 4$ is the postshock density compression). 
Importantly, $\zeta_{\rm tot}/4\approx 9/16$ suggests that the sum of all energy densities downstream is consistent with the one predicted in a high Mach number hydrodynamic shock.
\subsubsection{Acceleration Efficiency for Nonthermal Particles}
\label{subsec:epsilon}
%
Figures \ref{fig:epsilon}(a)-(d) show the acceleration efficiency for the NT protons and electrons in red and blue colors respectively.
For most shock parameters the proton acceleration efficiency is $\epsilon_{\rm nt, i}\approx 10\%$, consistent with previous studies \citep[e.g.,][]{caprioli+14a, crumley+19}.
The electron acceleration efficiency, $\epsilon_{\rm nt, e}$, is approximately $\sim 2\%$, and its trends with Mach numbers are similar to the particle injection efficiency displayed in Figure \ref{fig:eta_energetic}.
Importantly, for a realistic mass ratio $\epsilon_{\rm nt, e}\sim 1\%$, which increases marginally as $m_{\rm R}$ decreases.
\subsubsection{Magnetic and Thermal Energy Densities}\label{subsec:etheB}
%
Other important quantities in energy partitioning at the shock are the self-generated postshock magnetic energy density, 
which directly controls the synchrotron emissivity, and the electron thermal energy, which controls the thermal bremsstrahlung.
These quantities are shown in Figure \ref{fig:epsilon}.

Figure \ref{fig:epsilon} shows the partitioning of energy density in thermal plasma (brown -- protons and cyan -- electrons) and magnetic fields (orange).
While the profile of thermal particles is rather flat in the downstream, the magnetic energy density peaks right behind the shock (e.g., see Figure \ref{fig:magtheta}), which means that the exact value of $\epsilon_{\rm B}$ may depend on the integration region;
for the present analysis we have averaged over a length of $1000\,d_{\rm i}$ behind the shock, where also electron and proton spectra are calculated.

Figure \ref{fig:epsilon}(a) shows that $\epsilon_{\rm B}$ (orange curves) does not depend much on the shock speed. 
This is expected because the postshock magnetic fields are directly influenced by the upstream magnetic fluctuations produced by the proton-driven streaming instability. 
For example, when the ratio of the upstream to the downstream magnetic fields is parameterized by $\mathcal{R}_{\rm B}$, the postshock magnetic energy density derived from Equation \ref{eq:Bampli} yields
\begin{eqnarray}\label{eq:postB}
    \epsilon_{\rm B}\equiv \frac{\delta B^2/8\pi}{e_{\rm tot}}\sim \frac{\mathcal{R}_{\rm B}^2}{2}\frac{n_{\rm nt}}{n_{\rm 0}}\left(\frac{\langle p_{\rm nt,x}\rangle}{m_{\rm i}v_{\rm sh}}\right)\frac{1}{\zeta_{\rm tot}}.
\end{eqnarray}
Since the terms on the right-hand side are generally less influenced by the shock speed, $\epsilon_{\rm B}$ remains unaffected.
Using the ratio of the NT to thermal proton density in the upstream, $n_{\rm nt}/n_{\rm 0}\lesssim 10^{-2}$ (see e.g., Figure \ref{fig:Bperpmi}), $\mathcal{R}_{\rm B}\approx 4$, $\langle p_{\rm nt,x}\rangle/m_{\rm i}v_{\rm sh}\approx 2$ (Equation \ref{eq:p0alpha}), and $\zeta_{\rm tot}\approx 2.3$, Equation \ref{eq:postB} returns $\epsilon_{\rm B}\sim 0.06$, accurately reproducing the order of magnitude of $\epsilon_{\rm B}$ observed in the orange curve of Figure \ref{fig:epsilon}(a).

Comparing $\epsilon_{\rm B}$ with $\epsilon_{\rm  nt,e}$ in Figure \ref{fig:epsilon}(a) (orange and blue curves), we find that $\epsilon_{\rm  nt, e}/\epsilon_{\rm B}$ decreases with the shock speed, a trend similar to the one reported by \citet{reynolds+21} for Galactic SNRs (see their figure 7).
Note that the rapid decrease in $\epsilon_{\rm nt, e}/\epsilon_{\rm B}$ for $v_{\rm pt}/c=0.2$ in Figure \ref{fig:epsilon}(a) is a result of the superluminal effect, which might be exaggerated in our one-dimensional shock simulations (see \S \ref{subsec:vp}).

While $\epsilon_{\rm B}$ is almost independent of the shock speed, it does depend on the Mach numbers, as shown in Figures \ref{fig:epsilon}(b) and (c).
The reduction in $\epsilon_{\rm B}$ with increasing \alfven Mach number is a consequence of the lower magnetization of the upstream plasma, though the dependence of $\epsilon_{\rm B}$ on $\mathcal{M}_{\rm A}$ becomes weak when the nonresonant instability in the upstream is at work (Equations \ref{eq:Bampli} and \ref{eq:postB}), which typically occurs for $\mathcal{M}_{\rm A}\gtrsim 20$ \citep[see also][]{caprioli+14b,park+15}.
For a fixed $\mathcal{M}_{\rm A}$, instead, we notice that a warmer upstream (lower $\mathcal{M}_{\rm s}$) corresponds to a weaker magnetic field, likely due to the effectiveness of damping in higher-$\beta$ plasmas \citep[e.g.,][]{lee+73,volk+81,squire+17} (Figure \ref{fig:epsilon}(c), and also see Figure \ref{fig:magtheta}(c)).

When comparing the energy density of thermal protons $\epsilon_{\rm th,i}$ (shown by the brown curves) with that of other populations, we observe that a major fraction of the shock energy is processed into thermal protons. 
The thermal electron energy density $\epsilon_{\rm th,e}$ is only slightly lower than $\epsilon_{\rm th,i}$, suggesting that downstream thermal electrons have caught up with thermal protons, which we elaborate in the following section.
\subsection{Postshock Temperature}\label{subsec:TeTp}
%
\begin{figure*}
\centering
\includegraphics[width=5.5in]{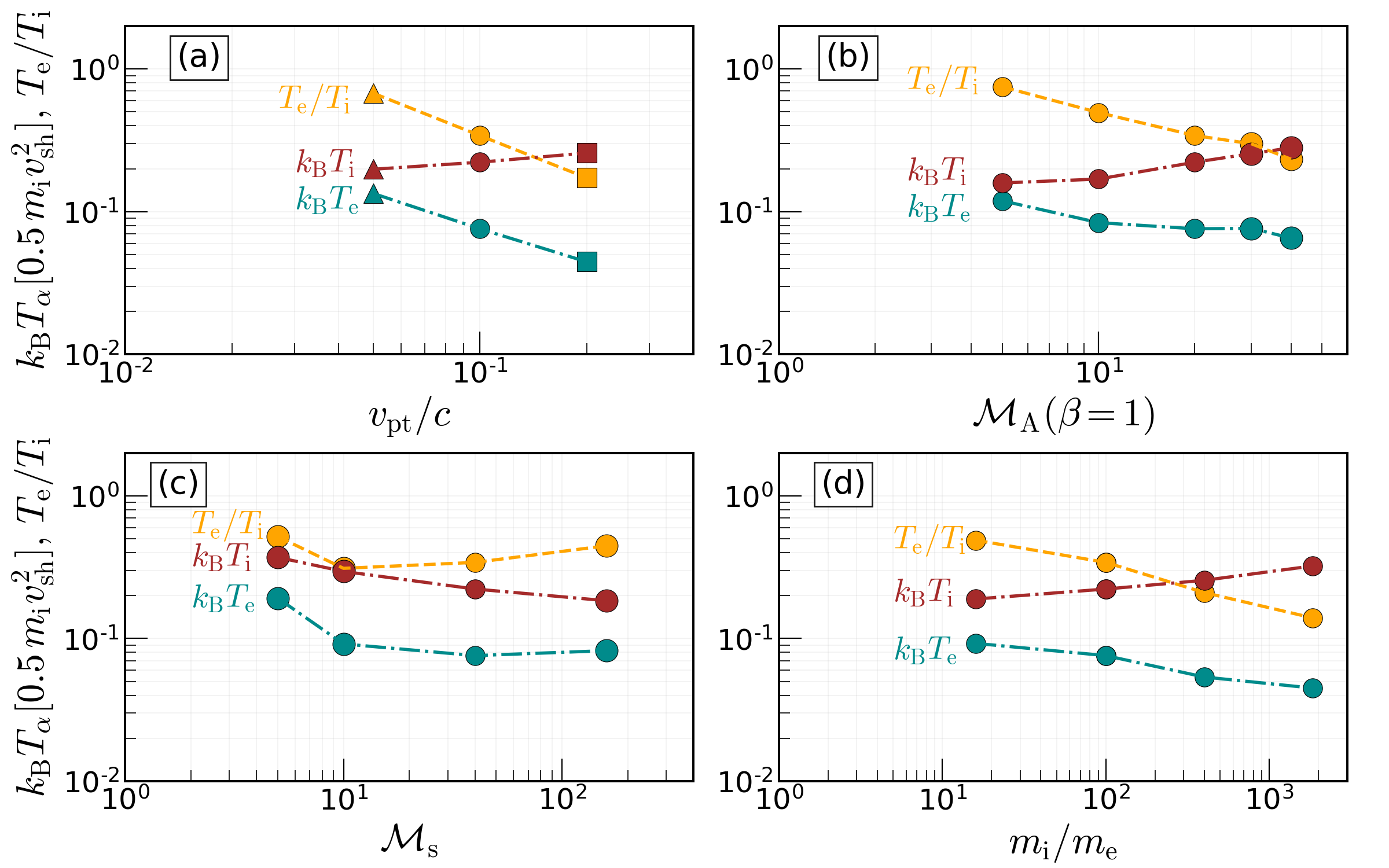}
\caption{The temperature of downstream thermal plasma for different shock parameters (Table \ref{tab:simpara}).
The brown and cyan lines show the temperature of protons and electrons normalized to $0.5m_{\rm i}v_{\rm sh}^2$ and the orange lines represent the electron-to-proton temperature ratio ($T_{\rm e}/T_{\rm i}$). 
}
\label{fig:tempds}
\end{figure*}
We now characterize the temperature of the downstream thermal plasma, comparing the results with the predictions of a collisionless system, where Coulomb interactions and inter-species energy exchanges are negligible.
In such a system, thermal electrons and protons are decoupled fluids: each species experiences its own shock jump and the downstream temperature for protons/electrons is given by
\begin{equation}\label{eq:Tds}
\frac{k_{\rm B } T_{\rm i,e}}{0.5\,m_{\rm i}v_{\rm sh}^2} =\frac{9}{8}\frac{m_{\rm i,e}}{m_{\rm i}} \frac{[1-\frac{(\gamma_{\rm th}-1)}{2\mathcal{M}^2_{\rm si,e}}][\frac{\gamma_{\rm th}}{\mathcal{M}^2_{\rm si,e}} + \frac{(\gamma_{\rm th}-1)}{2}] }{(1-\gamma_{\rm th}/\mathcal{M}^2_{\rm si,e})^2} \ ,
\end{equation}
where $\gamma_{\rm th}=5/3$ and $\mathcal{M}_{\rm si,e}=v_{\rm sh}/v_{\rm thi,e}$ is the ion/electron Mach number far upstream.

In the limit of strong shocks ($\mathcal{M}_{\rm si}\gg 1$), Equation \ref{eq:Tds} returns $k_{\rm B}T_{\rm i}/(0.5\,m_{\rm i}v_{\rm sh}^2)\simeq  0.375 $.
When comparing it with our simulations, shown by the brown curves in Figure \ref{fig:tempds}, we find that the proton temperature is smaller by a factor of $\lesssim 2$ for all parameters (see also Figure \ref{fig:shockstruc}).
Such differences are expected since the derivation of Equation \ref{eq:Tds} ignores energy transfer to electrons and magnetic fields.

In the case of electrons, the same choice $\mathcal{M}_{\rm se}\gg 1$ returns $k_{\rm B}T_{\rm e}/(0.5 m_{\rm i}v_{\rm sh}^2)\simeq  0.375/m_{\rm R}$, i.e., electron temperature is expected to be smaller than proton temperature by a factor of $m_{\rm R}$.
However, our analysis does not show much dependence on $m_{\rm R}$, see, e.g., the cyan curve in Figure \ref{fig:tempds}(d).
In fact, the electrons are roughly (within a factor of $\sim 3$) in equipartition with protons for all parameters (Figures \ref{fig:tempds}(a)-(d)), consistent with supernova remnants observations \citep[see, e.g.,][]{ghavamian+07}.
Moreover, in our simulations, $T_{\rm e}/T_{\rm i}$ is always found be much larger than $1/m_{\rm R}$.
The discrepancy is due to the non-adiabatic heating ensuing from the ion-driven instabilities in the precursor and in the shock foot, which control the electron temperature \citep{gupta+24a}.

To summarize, we show that the downstream thermal protons are cooler than that predicted in hydrodynamic quasi-parallel shocks and in general the final $T_{\rm e}/T_{\rm i}\sim 0.3$. 
The $T_{\rm e}/T_{\rm i}$ ratio decreases with increasing shock speed and $\mathcal{M}_{\rm A}$.
We observe a slight decrease in $T_{\rm e}/T_{\rm i}$ as $m_{\rm R}$ increases, while it rises with $\mathcal{M}_{\rm s}$, especially when $\mathcal{M}_{\rm se}\gtrsim 1$.
%

\section{Comparison with Previous Studies}\label{sec:compre}
Our 1D kinetic survey shows that in quasi-parallel non-relativistic shocks, electron injection efficiency $\eta_{\rm nt, e}\approx 0.5 \%$ and acceleration efficiency $\epsilon_{\rm nt, e}\approx2\%$.
The results qualitatively agree with the previous findings, where the NT efficiencies are estimated using a threshold of $p= 5 p_{\rm th}$ \citep[e.g.,][]{arbutina+21}.
Our $\eta_{\rm nt, e}$ and $\epsilon_{\rm nt, e}$ are slightly (factor of $\sim 2$) smaller than those reported for quasi-perpendicular shocks ($\theta_{\rm Bn}=63^{o}$) with a similar set of Mach numbers \citep{xu+20}.
The acceleration efficiency in our simulations attains an asymptotic value in $\mathcal{M}_{\rm A,s}$ for $\mathcal{M}_{\rm A,s}\gtrsim 20$.
Moreover, due to proton-driven electromagnetic turbulence, in quasi-parallel shocks the NT tail continues to grow with time;
our simulations cover more than two decades in energy, but the process is self-sustaining, so we expect quasi-parallel shocks to be able to accelerate relativistic electrons to the larger and larger energies necessary for both synchrotron radio and X-rays in astrophysical sources.

\citet{shalaby+22} suggested that electron injection and acceleration may be strongly dependent on the effectiveness of the ``intermediate-scale streaming instability", which should be active only when $\mathcal{M}_{\rm A}\lesssim \sqrt{m_{\rm R}}/4$ \citep{shalaby+21, shalaby+23}.
Our low-Mach number shocks, with $\mathcal{M}_{\rm A}=5$, are in this regime and our results align with the energy/momentum span of NT electrons reported in \citet{shalaby+22}. 
However, we do not observe a marked suppression in electron injection for larger mass ratios in high-$\mathcal{M}_{\rm A}$ shock.
We discussed how the differences in the shape of the NT tails in the case of $\mathcal{M}_{\rm A}=5$ shock for $m_{\rm R}=100$ and $m_{\rm R}=1836$ arise mainly from the normalization of the thermal electron peak position on the momentum axis, rather from the intrinsic injection/acceleration efficiency.
In fact, the cutoff energy of the NT spectra is comparable in both runs (see Figure \ref{fig:spec-mr}).
Our simulations of higher-Mach shocks with $\mathcal{M}_{\rm A}=20>\sqrt{m_{\rm R}}/4$ find an acceleration efficiency a factor of $\gg 10$ larger than that reported by \citet{shalaby+22} (see their figure 4), which suggests that the threshold for the intermediate-scale instability may not be a threshold for electron injection. 
A possible reason behind this discrepancy may lie in the selection of different initial parameters: $\theta_{\rm Bn}=0^{\rm o}$ (strictly parallel) and the sonic Mach number $\mathcal{M}_{\rm s}=v_{\rm sh}/v_{\rm th}\sim 600$ (note we redefine the value of the mentioned $\mathcal{M}_{\rm s}$ using thermal speed). 
In our case, these parameters are $\theta_{\rm Bn}=30^{\rm o}$ and $\mathcal{M}_{\rm s}=40$, respectively (see the run $\mathcal{B}1_{\rm IV}$ in Table \ref{tab:simpara}). 
However, such a significant difference is unexpected once proton acceleration begins, as they reorganize the initial orientation of the magnetic field through streaming instabilities (see Figure \ref{fig:shockstruc}(b)) and increase the plasma pressure/temperature (Figure \ref{fig:shockstruc}(e)), which can reduce the effective sonic Mach number, especially for electrons \citep[][]{gupta+24a}.
A shock with $\mathcal{M}_{\rm s}\sim 600$ and $\mathcal{M}_{\rm A}=20$ represents a regime of $\beta\equiv 4(\mathcal{M}_{\rm A}/\mathcal{M}_{\rm s})^2\lesssim 0.004$ (strongly magnetized), while we consider here $\beta = 0.06-64$ typical of shocks propagating in the solar wind or in the interstellar/intracluster medium.

When comparing our benchmark run with the results of \citet{park+15}, we find that our $\eta_{\rm nt, e}$ is slightly larger by a factor of $\lesssim 2$.
These simulations have identical shock parameters, but are performed in different frames (downstream frame for \citet{park+15}, upstream frame here);
it is likely that the choice of the upstream frame is less prone to the noise produced by drifting plasmas and hence more accurate for capturing the growth of the self-generated waves that scatter reflected particles.
A common ground of agreement is that high-Mach number quasi-parallel shocks are capable of accelerating electrons and protons up to the maximum energies allowed by the system age/size, or synchrotron losses for the electrons.
As particle injection efficiency positively correlates with shock reflectivity, our results also support the notion that quasi-parallel shocks are effective in electron reflection \citep[see e.g.,][]{moris+22,bohdan22} — a topic we will discuss in an upcoming work (Gupta et al, in prep).

Our finding is consistent, for instance, with the radio/X-ray/$\gamma-$ray observations of SN1006 \citep[][]{SN1006HESS, giuffrida+22}, whose shock speed is $v_{\rm sh}\approx 3000\, {\rm km\,s^{-1}}$ and $\mathcal{M}_{\rm A}\sim 50$, assuming an interstellar magnetic field of $\sim 3{\rm \mu G}$ field and the ambient plasma density of $\sim 0.01\,{\rm cm^{-3}}$.
Notably, synchrotron $X$-rays detected in quasi-parallel regions of SN1006 independently confirm our results \citep[see e.g.,][]{winner+20}.

Finally, we compare our results with PIC simulations of  trans-relativistic and relativistic quasi-parallel shocks \citep[e.g.,][]{sironi+11,crumley+19}. 
In this regime particle injection is hindered by superluminal regions, which means that the initial  $\theta_{\rm Bn}$ and the generation of non-linear magnetic fluctuations are crucial for electron acceleration.
Relativistic subluminal shocks show $\eta_{\rm nt, e}\approx 2 \%$ and $\epsilon_{\rm nt, e}\approx10\%$, which are almost a factor of $\sim 4$ larger than non-relativistic shocks as reported in our work, which suggests that electron acceleration scales with the shock speed, provided that the shock never becomes superluminal.
The regime of weakly-magnetized shocks, where the initial direction of the upstream magnetic field becomes less important \citep{sironi+11, orusa+23, grassi+23}, deserves a dedicated treatment in the non-relativistic regime, which will be the subject of a future study.

Moreover, an open issue that remains is how the injection and acceleration of NT particles get modified in realistic environments where NT particles span several orders of magnitude in energy. 
The particle distribution following $f(p)\propto p^{-4}$ can lead to NT energy budget (i.e., acceleration efficiency) increasing with the maximum energy of NT particles unless the injection fraction is self-regulated and/or efficient acceleration induces steeper spectra \citep[as found in hybrid simulations, see e.g.,][]{caprioli+20}. 
The long-term regulation of injection and acceleration efficiency remains an open question and will be explored in future work.
\section{Conclusions}\label{sec:conclusion}
%
We have used fully kinetic 1D particle-in-cell simulations to study electron acceleration at quasi-parallel non-relativistic shocks for different shock speeds, and Alfv\'{e}n and sonic Mach numbers.
The key takeaways are:
\begin{enumerate}
 \setlength\itemsep{-0.005\textwidth}
     \item 
     {\it Acceleration efficiency:} For all shock parameters, we notice the development of a power-law tail for both electrons and protons, suggesting that quasi-parallel shocks can produce nonthermal (NT) particles.
     In the downstream, the NT tail is found at $p_{\rm i,e}\gtrsim 3\,m _{\rm i}v_{\rm pt}\sqrt{m_{\rm i,e}/m_{\rm i}}$ (where $m_{\rm i}$, $m_{\rm e}$, and $v_{\rm pt}$ denote proton mass, electron mass, and piston speed).
     The nonthermal electron and proton fractions in the downstream are $\sim 0.5\%$ and $\sim 1\%$, respectively. 
     The acceleration efficiency is approximately $2\%$ for electrons and $10\%$ for protons (Figure \ref{fig:epsilon}).
     \item {\it Self-generated magnetic fields and shock inclination}:  
     The streaming instability driven by energetic protons in the upstream is effective in reorienting and amplifying the initial magnetic field (Figure \ref{fig:shockstruc}(b)).
     The magnetic field amplification strongly depends on the \alfven Mach number (Equation \ref{eq:Bampli} and Figure \ref{fig:magtheta}).
    \item {\it Shock speed}: While acceleration is moderately faster in high-speed shocks (Figure \ref{fig:spec_vp}), 
    for $v_{\rm sh}/c\gtrsim 0.2$, the power-law tail stops growing beyond $\sim 10\, m_{\rm i}v_{\rm pt}$.
    This is because the self-generated magnetic field creates patches of superluminal configurations, which may however be exaggerated by the 1D setup (Figure \ref{fig:magtheta}(a)).
    For $v_{\rm sh}/c\lesssim 0.1$, the spectra are very similar and the upstream shows an extended precursor for electrons as well as for protons (Figures \ref{fig:x-vx-vp}).
    \item {\it \alfven Mach number}: 
    Large $\mathcal{M}_{\rm A}$ is preferred for acceleration, as it generates large amplitude upstream magnetic turbulence ($\delta B/B\gg 1$), which increases the confinement of particles and allows them to re-approach the shock multiple times. 
    For $\mathcal{M}_{\rm A}\lesssim 10$, acceleration stalls due to the lack of self-generated turbulence, at least over the timescales that we were able to probe (Figures \ref{fig:spec-Ma-Ms-MaMs}(a1)--(a3)).
    \item {\it Sonic Mach number}: 
    As the self-generated turbulence becomes weaker with increasing upstream plasma temperature, the acceleration efficiency decreases for smaller $\mathcal{M}_{\rm s}$ (Figure \ref{fig:epsilon}(c)). 
    However, unlike at low $\mathcal{M}_{\rm A}$, the non-thermal tail grows at a smaller rate without stalling (Figure \ref{fig:spec-Ma-Ms-MaMs}(b1)--(b4)).
    \item {\it Proton-to-electron mass ratio}:
    The NT fraction of electrons and protons does not depend much on the assumed mass ratio (Figure \ref{fig:eta_energetic}(d)), but since the thermal peaks are further away the electron-to-proton ratio $K_{\rm ei}$ measured at NT momenta decreases as $m_{\rm R}$ increases (Figure \ref{fig:spec-mr}). 
    This is reasonable since electrons have to go through more Fermi cycles and in each cycle some electrons are advected downstream.
    \item {\it Thermal and magnetic energy budgets}:
    We analyze how the downstream energy is distributed among thermal and NT particles, as well as magnetic fields (Figure \ref{fig:epsilon}).   
    The thermal energy density of electrons is smaller than that of protons by a factor $\sim 2$.
    For $\mathcal{M}_{\rm A}\gtrsim 20$ shocks, the downstream magnetic field is approximately $3\%$ of the total postshock energy density. 
    \item {\it Downstream temperature ratio}: 
    The electron-to-proton temperature ratio in the downstream thermal plasma depends mostly on shock speed and $\mathcal{M}_{\rm A}$.
    For $\mathcal{M}_{\rm A}\gtrsim 20$, $T_{\rm e}/T_{\rm i}\sim 0.5$, which becomes smaller by a factor of $\lesssim 2$ with increasing mass ratio (Figure \ref{fig:tempds}).
\end{enumerate}

In summary, our first-principles kinetic simulations offer detailed insights into electron acceleration and various observational parameters in quasi-parallel non-relativistic shocks. 
In a forthcoming work, we will use these results to build a comprehensive theoretical model of electron acceleration.

\section*{ACKNOWLEDGMENTS}
We thank the anonymous referee for reviewing the manuscript carefully and providing constructive suggestions.
We thank Christoph Pfrommer and Mohamad Shalaby for their insightful discussions.
DC was partially supported by NASA through grants 80NSSC20K1273 and 80NSSC18K1218 and NSF through grants PHY-2010240 and AST-2009326. 
AS acknowledges the support of NSF grant PHY-2206607 and the Simons Foundation grant MP-SCMPS-00001470.
Simulations were performed using the computational resources provided by the University of Chicago Research Computing Center and ACCESS (TG-AST180008).
\appendix
\section{Evolution of the magnetic field in different mass ratio runs}\label{app:Bfieldmi}
\begin{figure}
    \centering
\begin{minipage}{.4\linewidth}
\centering
\includegraphics[width=2.4in]{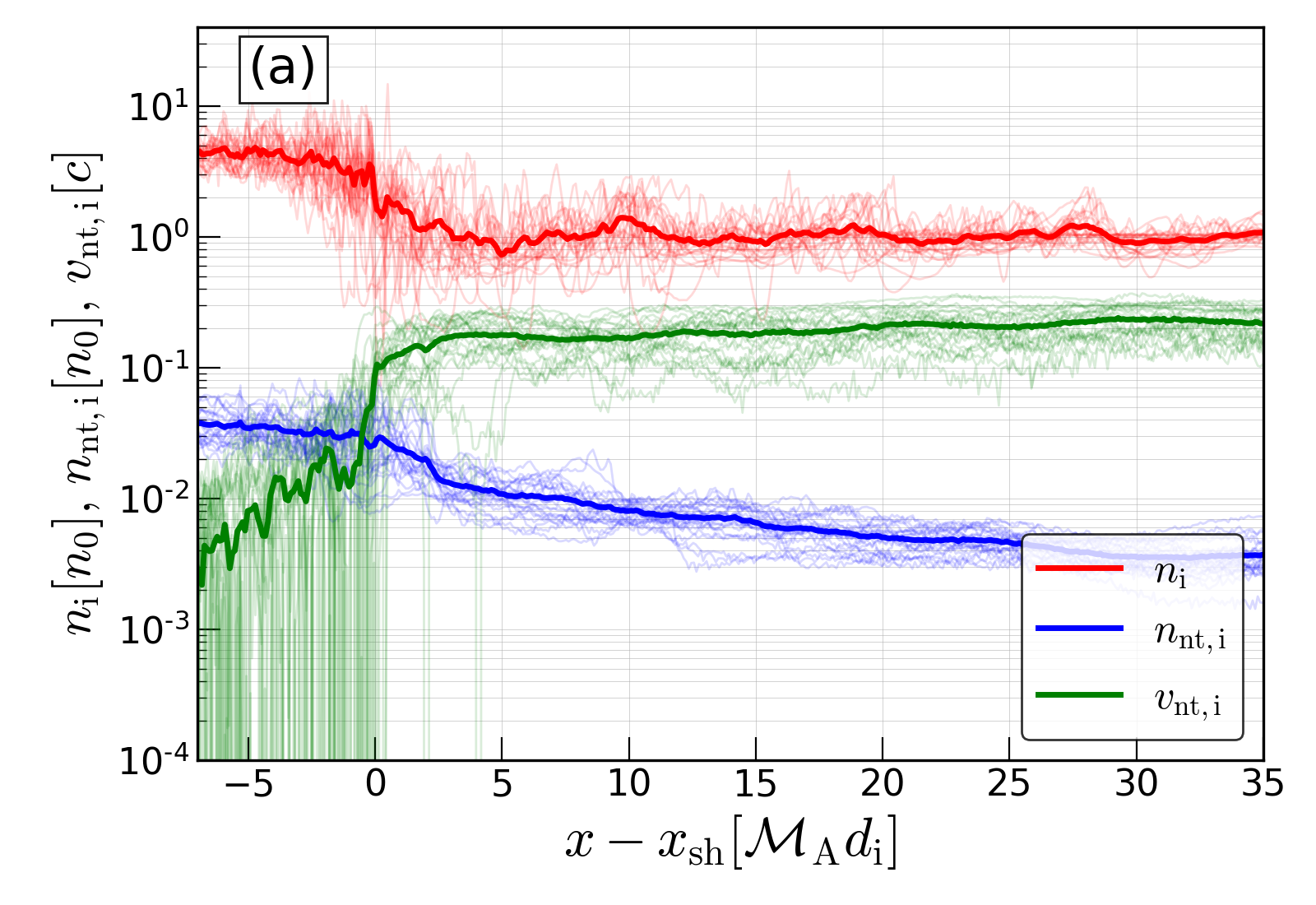}
\end{minipage}
\hspace{0.05\linewidth}
\begin{minipage}{.4\linewidth}
\centering
\includegraphics[width=2.4in]{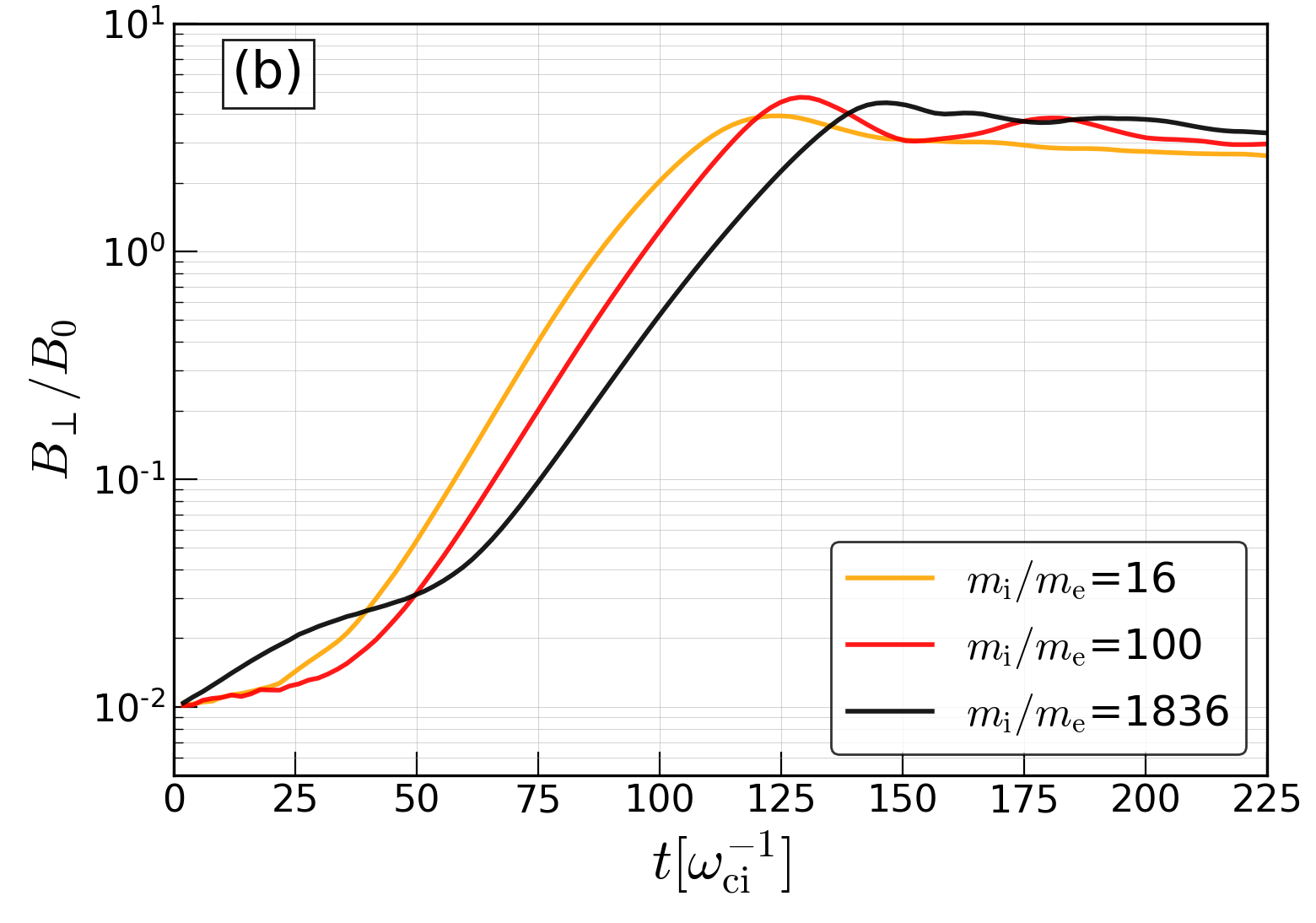}
\end{minipage}
    \caption{
    Panel (a) - diagnostics showing the beam plasma parameters in our benchmark shock run $\mathcal{B}1$ in Table \ref{tab:simpara} and panel (b) - the time evolution of $B_{\rm \perp}/B_{\rm 0}$ for the three controlled simulations using the parameters found in the shock run of the panel (a).
    Thick lines in panel (a) show the time-averaged (between $100-150\,\omega_{\rm ci}^{-1}$) profile of thermal and NT proton number densities (normalized to far upstream thermal proton density $n_{\rm 0}$), and the drift velocity of NT protons ($v_{\rm nt,i}$) in the background plasma rest frame.
    Note in panel (b), how for larger $m_{\rm R}$ the growth is delayed, though eventually the same saturation is achieved.}
    \label{fig:Bperpmi}
\end{figure}

Here, we present a diagnostic of the NT proton beam that drives streaming instability and produces large amplitude magnetic fields upstream of our benchmark shock simulation (the run $\mathcal{B}1$ in Table \ref{tab:simpara}).
We use these parameters in our controlled simulations to investigate the dependence of the self-generated magnetic field on $m_{\rm R}$, as found in Figure \ref{fig:magtheta}(f). 
Figure \ref{fig:Bperpmi}(a) shows the NT proton beam in our benchmark shock run drifting at a speed of $v_{\rm nt,i} \equiv 2v_{\rm pt} = 0.2\,c$ relative to the upstream thermal plasma, with the beam number density of $n_{\rm nt,i}= 0.01 n_{\rm 0}$.
Using these parameters in the nonresonant streaming instability setup \citep{gupta+21}, we perform three periodic-box controlled simulations, where each simulation differs only by $m_{\rm R}$, while keeping thermal background plasma parameters identical to those in our benchmark shock simulation, i.e., $v_{\rm th,i} = 3.34 \times 10^{-3}\,c$ and $v_{\rm A} = 6.67 \times 10^{-3}\,c$. 
The time evolution of the total $B_{\rm \perp}$ for these controlled simulations is shown in Figure \ref{fig:Bperpmi}(b).

We find that our controlled simulations produce $B_{\rm \perp}/B_{\rm 0} > 1$, as expected from the nonresonant streaming instability \citep{bell04,amato+09}.
Although the magnetic field at saturation is similar for different $m_{\rm R}$, the onset of linear growth is delayed for larger mass ratios. 
These results suggest that unless we perform time averages in the plateau region, which occurs at $t > 150\,\omega_{\rm ci}^{-1}$ for our benchmark run, $B_{\rm \perp}/B_{\rm 0}$ is likely to decrease with increasing $m_{\rm R}$, as observed in our shock simulations (Figure \ref{fig:magtheta}(f)). 
Additionally, in shock simulations, the continuous evolution of NT particles' anisotropic momenta, which determine the amplification of the magnetic field, makes a one-to-one comparison of different mass ratio simulations extremely challenging.

\section{Characterizing streaming instabilities in the upstream of a shock}\label{app:resonantvsnonresonant}
To sustain DSA, NT particles need to be confined close to the shock, which requires electromagnetic fluctuations present in the upstream. 
While both resonant and nonresonant streaming instabilities can contribute to the injection and acceleration of particles, nonresonant instability is advantageous due to its ability to produce $\delta B/B_{\rm 0}\gg 1$. 
For nonresonant streaming instability to occur, two conditions must be satisfied: (1) $\gamma_{\rm fast} < \omega_{\rm ci}$ and (2) $\mathcal{P}_{\rm nt,i}\gg 2P_{\rm B,0}$, where $\gamma_{\rm fast}$ is the growth rate of the fastest-growing mode, $\omega_{\rm ci}$ is the gyrofrequency of the background thermal ions, $\mathcal{P}_{\rm nt,i}$ is the anisotropic momentum of NT particles that drive the instability, and $P_{\rm B,0}$ is the initial magnetic pressure in the upstream plasma \citep[][]{bell04,amato+09,gupta+21,zacharegkas+24}.  
The first condition ensures that the instability evolves more slowly than the plasma response. 
The second condition requires that the motion of the instability-driving particles remains largely unaffected in the linear regime of the instability, i.e., deflection of the driving protons occurs only when the magnetic fields evolve into the nonlinear regime ($\delta B/B_{\rm 0} \gtrsim 1$).
From previous numerical simulations, we find that $\mathcal{P}_{\rm nt,i}/2 P_{\rm B,0}\gtrsim 10\equiv \xi_{\rm min}$ is needed to clearly identify the nonresonant streaming instability (see figure 9 in \citealt{gupta+21}).
Using the above conditions, below we estimate a possible range of Mach numbers for the nonresonant instability.

As the growth rate of nonresonant instability is given by $\gamma_{\rm fast} = 0.5 (n_{\rm nt,i}/n_{\rm 0}) (v_{\rm d}/v_{\rm A0}) \omega_{\rm ci}\equiv 0.5 J_{\rm nt,i}/(n_{\rm 0} e v_{\rm A0})$ \citep[e.g.,][]{bell04}, the condition (1) yields $J_{\rm nt,i}/(n_{\rm 0} e v_{\rm A0})<2$.
In the shock simulations, we usually find that in the upstream $n_{\rm nt,i}/n_{\rm 0} \lesssim 0.01$ and $v_{\rm d} \sim 2\,v_{\rm pt}$ (for example, see Figure \ref{fig:Bperpmi}(a)). 
This sets the upper limit of the Alfv\'{e}n Mach number for the nonresonant instability, which is $\mathcal{M}_{\rm A} \lesssim 2/(n_{\rm nt,i}/n_{\rm 0}) \approx 200$. 
For shocks with $\mathcal{M}_{\rm A} \gg 200$, a Weibel-type instability may be seen -- a regime that is not addressed in this work.
Similarly, the condition (2) gives $\mathcal{M}_{\rm A} \gtrsim \sqrt{\xi_{\rm min}/(4\,n_{\rm nt,i}/n_{\rm 0})} \approx 15$, where we take $\mathcal{P}_{\rm nt,i} \approx n_{\rm nt,i} \langle p_{\rm nt,x} \rangle v_{\rm d}$, with the average $x$-momentum of NT protons assumed to be $\langle p_{\rm nt,x} \rangle \gtrsim 2 m_{\rm i} v_{\rm sh}$. 
Note that the lower limit may be influenced by the maximum energy of NT particles, as $\mathcal{P}_{\rm nt,i}$ can increase with the growth of the NT tail in higher energies.

\begin{figure}
\centering
\begin{minipage}{.45\linewidth}
\centering
\includegraphics[width=3.4in]{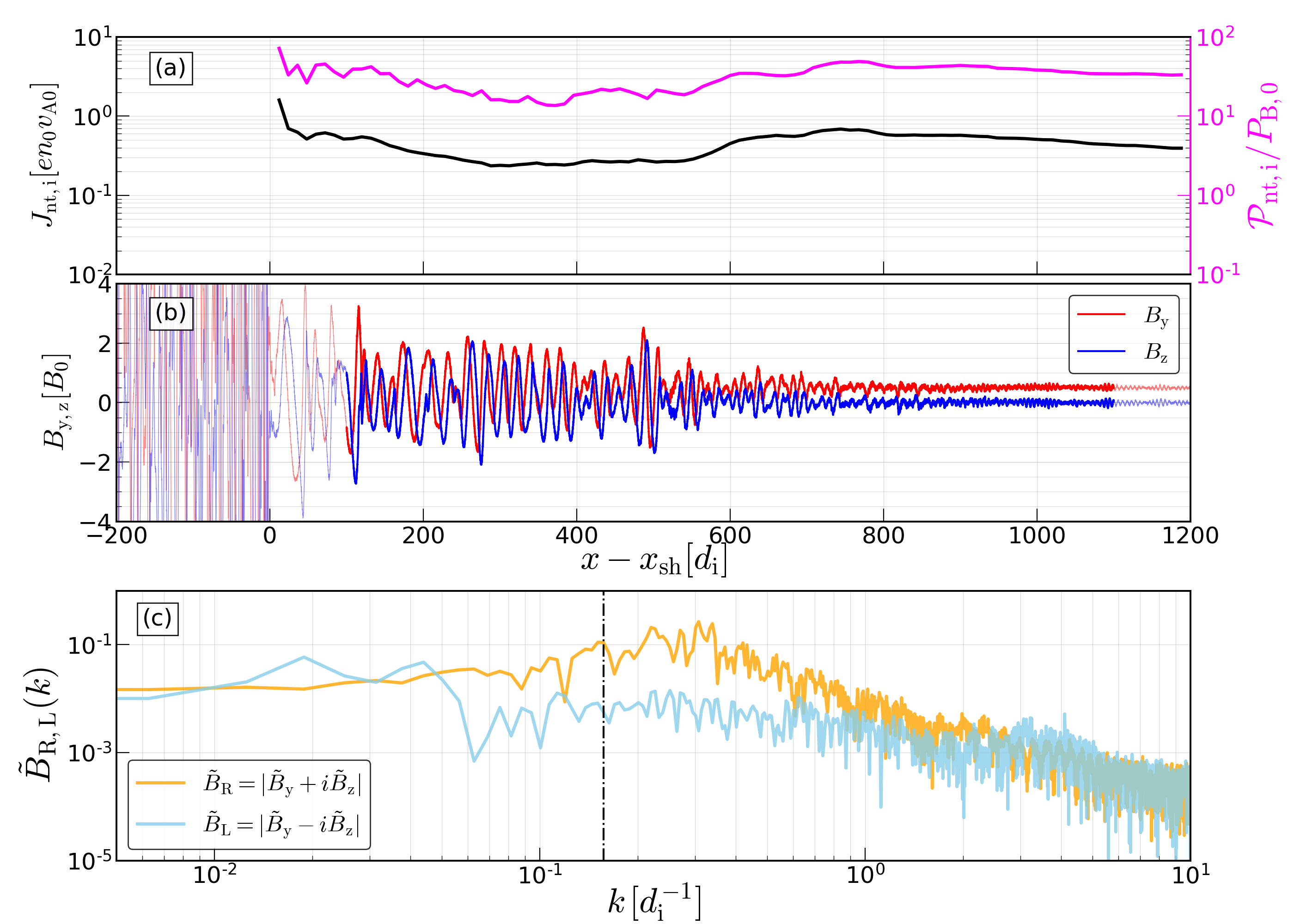}
\end{minipage}
\hspace{0.029\linewidth}
\begin{minipage}{.45\linewidth}
\centering
\includegraphics[width=3.4in]{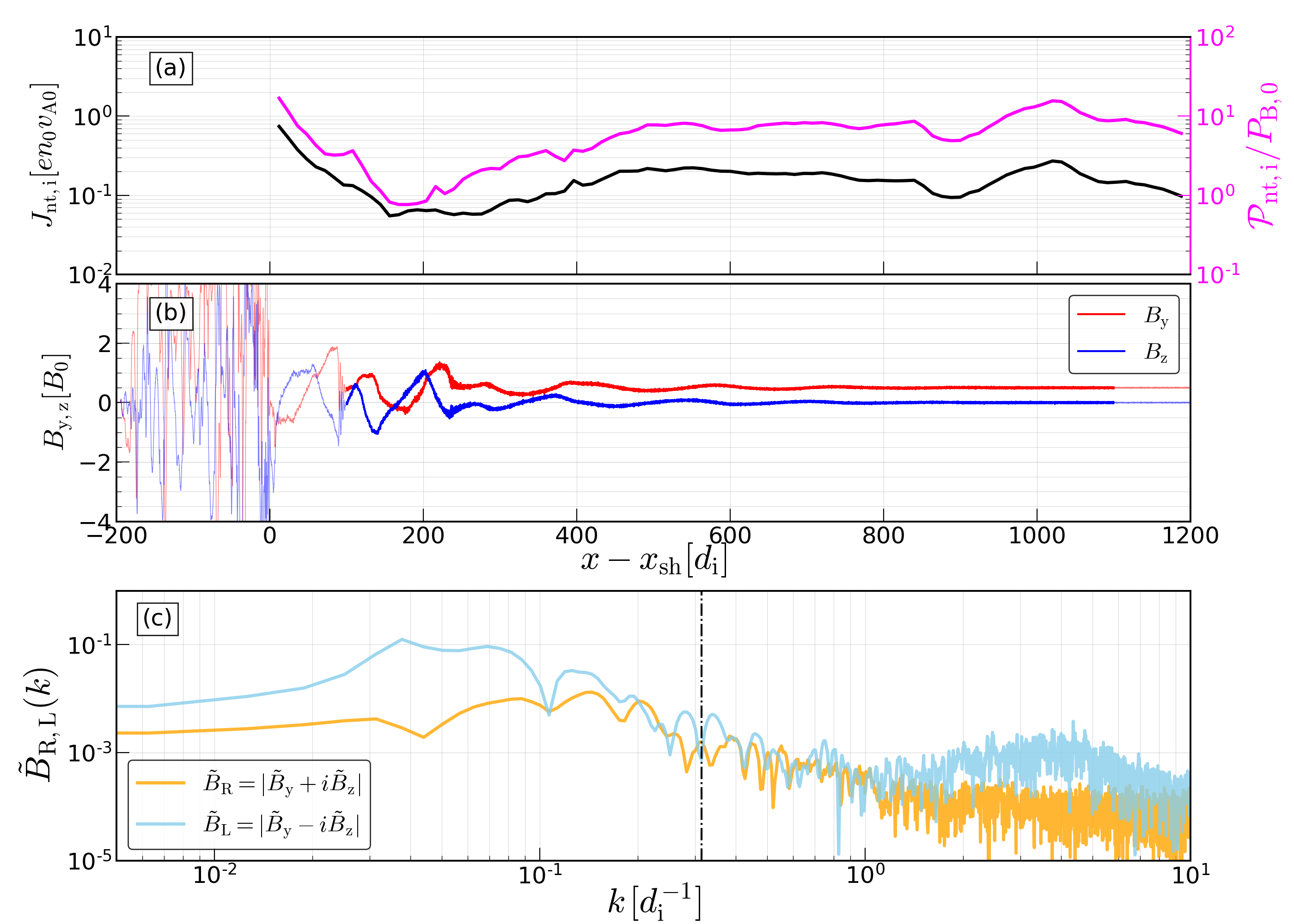}
\end{minipage}
   \caption{
    Investigating the nature of streaming instability in our shock simulations.
    The left and right panels correspond to two different shock simulations, $\mathcal{M}_{\rm A}=20$ (the run $\mathcal{B}1$) and $\mathcal{M}_{\rm A}=10$ (the run $\mathcal{B}3$), respectively at $t\approx 95\,\omega_{\rm ci}^{-1}$, where all other parameters $v_{\rm pt}/c=0.1$, $\mathcal{M}_{\rm s}=40$, and $m_{\rm i}/m_{\rm e}=100$ are identical. 
    Panel (a) shows the current density (black) and anisotropic pressure (magenta) of NT protons in the upstream, panel (b) shows the profiles of the magnetic fields, and panel (c) represents the Fourier mode analyses of the magnetic fields for the thick blue/red region of the panel (b).
    In panel (c), the vertical dashed black line marks the resonant wavenumber, $k=2\pi/R_{\rm L}\equiv \pi/\mathcal{M}_{\rm A}\,d_{\rm i}^{-1}$, for particles propagating at twice the shock speed. The panels on the left suggest that $\mathcal{P}_{\rm nt,i}/P_{\rm B,0}\gtrsim 20$ for the $\mathcal{M}_{\rm A}=20$ shock, resulting in predominantly right-handed (nonresonant) magnetic fluctuations, $\tilde{B}_{\rm R}$. 
    In contrast, for the $\mathcal{M}_{\rm A}=10$ shock in the right panels, the upstream fluctuations are mostly left-handed $\tilde{B}_{\rm R}$, indicating that they are resonant with the current-driving protons.
    Note that this analysis applies to the linear growth stage of the instability. 
    At later times, both resonant and nonresonant waves may exhibit equal power in the shock precursor (where NT particles diffuse), and the linear growth shifts to regions further upstream where NT particles stream.
    }
    \label{fig:Binstb}
\end{figure}

To summarize, in our shock simulations, the nonresonant instability is found for $\mathcal{M}_{\rm A} = 20$, $30$, and $40$, while the resonant instability is observed for $\mathcal{M}_{\rm A} = 5$ and $10$.
Resonant and nonresonant modes have different helicities and can be directly identified from the profiles of the magnetic fields, as illustrated in Figure \ref{fig:Binstb}.

\bibliography{Total}{}
\bibliographystyle{aasjournal}

\end{document}